\documentclass[aps,prl,twocolumn,preprintnumbers,superscriptaddress,dblfloatfix,nofootinbib]{revtex4-1}
\usepackage[utf8]{inputenc}
\usepackage{lmodern}
\usepackage{amsmath,amssymb,mhchem}
\usepackage{graphicx}
\usepackage{url}
\usepackage{color}
\usepackage{enumitem}
\usepackage{subfigure}
\usepackage[dvipsnames]{xcolor}
\usepackage[colorlinks=true,breaklinks=true]{hyperref}
\hypersetup{allcolors=[rgb]{0.0 0.0 0.6},linkcolor=[rgb]{0.75 0.05 0.05}}
\usepackage{bm}
\usepackage{epsfig}
\usepackage{amsmath}
\usepackage{amssymb}
\usepackage{slashed}
\usepackage{color}
\usepackage{accents}
\usepackage[dvipsnames]{xcolor}
\usepackage[colorlinks=true,breaklinks=true]{hyperref}
\hypersetup{allcolors=[rgb]{0.0 0.0 0.6},linkcolor=[rgb]{0.75 0.05 0.05}}

\newcommand{\exclude}[1]{}

\begin{document}

\preprint{IPMU20-0006}
\preprint{YITP-20-11 \ \   }

\title{Exploring Primordial Black Holes from the Multiverse with Optical Telescopes}

\author{Alexander Kusenko}
\affiliation{Department of Physics and Astronomy, University of California, Los Angeles \\ Los Angeles, California, 90095-1547, USA}
\affiliation{Kavli Institute for the Physics and Mathematics of the Universe (WPI), UTIAS \\The University of Tokyo, Kashiwa, Chiba 277-8583, Japan}

\author{Misao Sasaki}
\affiliation{Kavli Institute for the Physics and Mathematics of the Universe (WPI), UTIAS \\The University of Tokyo, Kashiwa, Chiba 277-8583, Japan}
\affiliation{Center for Gravitational Physics, Yukawa Institute for
Theoretical Physics, \\
Kyoto University, Kyoto 606-8502, Japan}
\affiliation{Leung Center for Cosmology and Particle Astrophysics,
National Taiwan University, 
Taipei 10617, Taiwan}

\author{Sunao Sugiyama}
\affiliation{Kavli Institute for the Physics and Mathematics of the Universe (WPI), UTIAS \\The University of Tokyo, Kashiwa, Chiba 277-8583, Japan}
\affiliation{Department of Physics, The University of Tokyo, 7-3-1 Hongo, Bunkyo-ku, Tokyo 113-0033 Japan}

\author{Masahiro Takada}
\affiliation{Kavli Institute for the Physics and Mathematics of the Universe (WPI), UTIAS \\The University of Tokyo, Kashiwa, Chiba 277-8583, Japan}

\author{Volodymyr Takhistov}
\affiliation{Department of Physics and Astronomy, University of California, Los Angeles \\ Los Angeles, California, 90095-1547, USA}

\author{Edoardo Vitagliano}
\affiliation{Department of Physics and Astronomy, University of California, Los Angeles \\ Los Angeles, California, 90095-1547, USA}

\date{\today}

\begin{abstract}
Primordial black holes (PBHs) are a viable candidate for dark matter if the PBH masses are in the currently unconstrained ``sublunar'' mass range. We revisit the possibility that PBHs were produced by nucleation of false  vacuum bubbles during inflation. We show that this scenario can produce a population of PBHs that simultaneously accounts for all dark matter, explains the candidate event in Subaru Hyper Suprime-Cam (HSC) data, and contains both  heavy black holes as observed by LIGO and  very heavy seeds of supermassive black holes. We demonstrate with numerical studies that future observations of HSC, as well as other optical surveys, such as LSST, will be able to provide a definitive test for this generic PBH formation mechanism if it is the dominant source of dark matter. 
\end{abstract}
\maketitle

Primordial black holes (PBHs), formed in the early Universe prior to any galaxies and stars, are a viable candidate for dark matter~(e.g.~\cite{Zeldovich:1967,Hawking:1971ei,Carr:1974nx,GarciaBellido:1996qt,Khlopov:2008qy,Frampton:2010sw,Kawasaki:2016pql,Cotner:2016cvr,Cotner:2017tir,Carr:2016drx,Inomata:2016rbd,Pi:2017gih,Inomata:2017okj,Garcia-Bellido:2017aan,Inoue:2017csr,Georg:2017mqk,Inomata:2017vxo,Kocsis:2017yty,Ando:2017veq,Cotner:2016cvr,Cotner:2017tir,Cotner:2018vug,Sasaki:2018dmp,Carr:2018rid,Cotner:2019ykd}). It has also been suggested that they could play a central role in a variety of astrophysical phenomena, such as 
progenitors~\cite{Nakamura:1997sm,Clesse:2015wea,Bird:2016dcv,Raidal:2017mfl,Eroshenko:2016hmn,Sasaki:2016jop,Clesse:2016ajp,Fuller:2017uyd}
for the LIGO gravitational wave events~\cite{Abbott:2016blz,Abbott:2016nmj,Abbott:2017vtc}, seeds for formation of supermassive black holes~\cite{Bean:2002kx,Kawasaki:2012kn,Clesse:2015wea} as well as the source of new signals~\cite{Fuller:2017uyd,Takhistov:2017nmt,Takhistov:2017bpt} from compact star disruptions from PBH capture, among others. 

PBHs can form through a variety of mechanisms (see e.g.~\cite{Khlopov:2008qy,Carr:2016drx} for review). While many models focus on inflationary perturbations as a source of PBHs, other formation mechanisms, such as cosmic string collapse~\cite{Caldwell:1995fu,Garriga:1992nm}, bubble collisions~\cite{Hawking:1982ga,Lewicki:2019gmv}, domain wall collapse~\cite{Garriga:1992nm,Khlopov:2008qy,Deng:2016vzb} as well as scalar field fragmentation~\cite{Cotner:2016cvr,Cotner:2018vug,Cotner:2019ykd} can produce copious populations of PBHs. Depending on the formation time, resulting PBHs can span many orders of magnitude in mass. Those formed with mass above the Hawking evaporation limit of $\sim 10^{15}$~g survive will survive until the present day. The abundance of PBHs with larger masses have been constrained with astrophysical observations.~On the other hand, recent reanalyses~\cite{Katz:2018zrn,Smyth:2019whb,Montero-Camacho:2019jte} of PBHs in the lower ``sublunar'' mass range range of $\sim 10^{-16} - 10^{-10} M_{\odot}$
have established that there remains a sizable open parameter space window for PBHs to constitute all of the dark matter.

In this work we revisit a generic scenario of PBH formation from vacuum bubble nucleation during inflation~\cite{Garriga:2015fdk,Deng:2017uwc,Deng:2018cxb}. We will show that the resulting broad mass function of PBHs can simultaneously account for all of the DM, the  
observed LIGO events, and also provide seeds for supermassive black holes  (SMBHS). Furthermore, a candidate event from the Subaru Hyper Suprime-Cam (HSC) microlensing search~\cite{Niikura:2017zjd} is consistent with this scenario. In particular, while the mass function of PBHs peaks at much smaller masses, where microlensing effect is negligible, the large-mass tail overlaps with the HSC sensitivity range, and it is consistent with detection of the reported candidate event~\cite{Niikura:2017zjd}. Upcoming HSC observations and other optical surveys will be able to test vacuum bubble formation as the primary source of DM in the form of PBH.
 
We assume that inflation took place in the early universe. The energy density of the inflaton field $\rho_i$ evolves slowly during the slow-roll phase of inflation. In addition to the inflaton and the experimentally discovered Higgs boson, other scalar fields are likely to exist. Such fields appear in a number of models of new physics, including supersymmetry and string theory~\cite{Susskind:2003kw}. This naturally leads one to consider multi-field potential for the inflaton. If the multi-field potential has a local minimum with energy density $\rho_b$  close to the path of the inflaton, there is a possibility of tunneling to it via Coleman -- De Luccia instanton~\cite{Coleman:1980aw}.  Let us consider the case $0< \rho_b < \rho_i$. During the slow-roll phase, the false vacuum can be populated repeatedly in a series of bubbles, each of which has energy density $\rho_b$ in the interior.  While these bubbles can expand, they do not percolate since the space outside the bubbles expands at a high rate. 

\begin{figure}[tb]
  \includegraphics[width=0.85\linewidth]{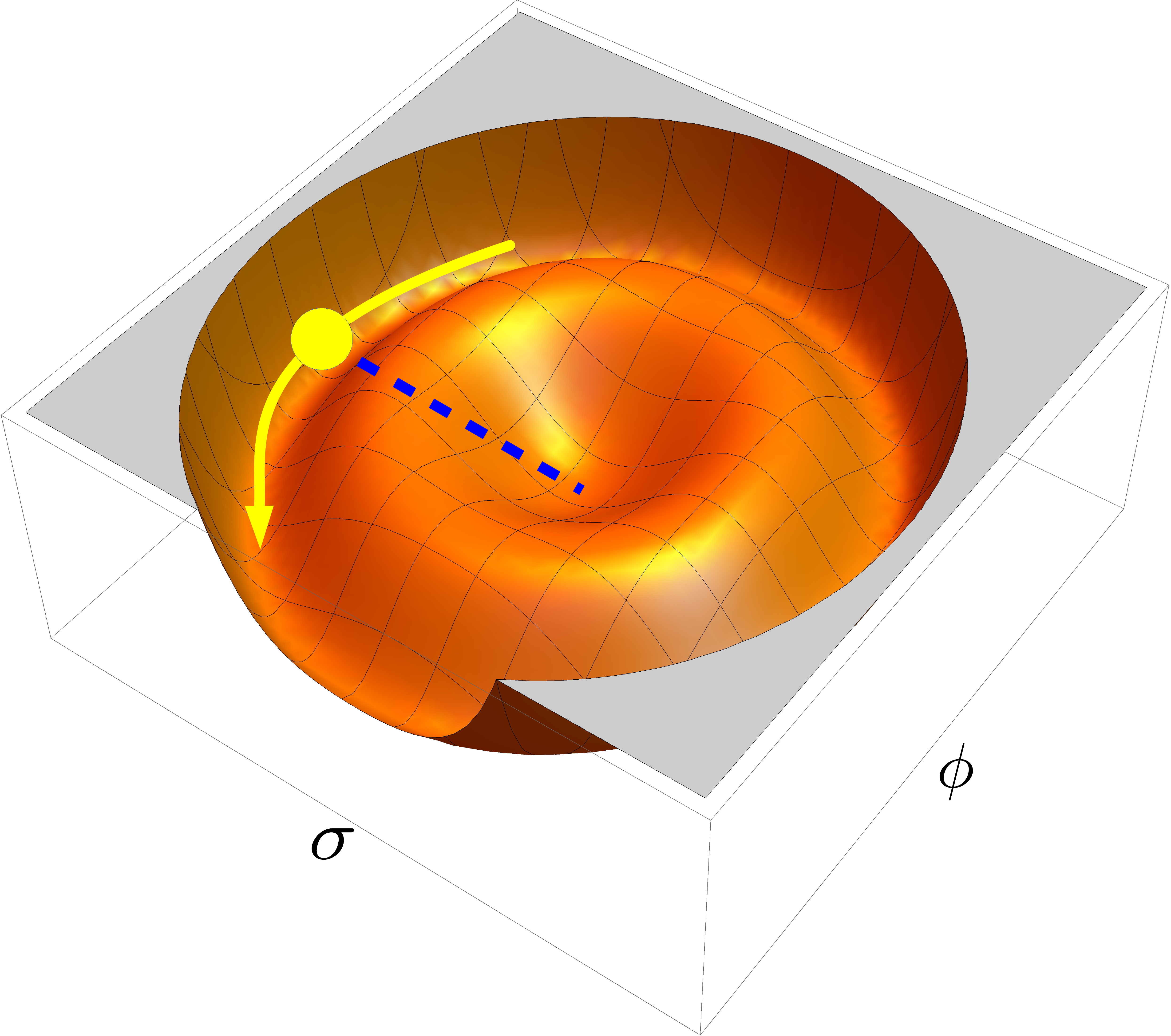}
\caption{Illustration of tilted ``Mexican hat'' potential $V(\phi, \sigma)$ describing a slowly rolling field tunneling to a minimum at the origin at an approximately constant rate.}
\label{fig:vacbubpot}
\end{figure}

Let us illustrate the qualitative features of a nearly constant bubble production over a period of slow-roll inflation using a 2-field potential of the type
\begin{align} \label{eq:model}
    V( \phi, \sigma ) =&~ m^2  ( \phi^2 + \sigma^2) - a ( \phi^2 + \sigma^2)^2 \\
    &~+ \dfrac{c}{M_{\rm pl}^2}  ( \phi^2 + \sigma^2)^3 + g M_{\rm pl}^4 \sin\Big(\dfrac{\phi}{f  M_{\rm pl}}\Big)~,\notag
\end{align}
where $g \ll m^2/M_{\rm Pl}^2 < c \sim a$ and $f > 1$. The potential, depicted in Fig.~\ref{fig:vacbubpot}, resembles a ``Mexican hat'' with a dent at the origin and a small tilt due to the shift-symmetric term  $\sin(\phi/f M_{\rm pl})$, which breaks the rotational symmetry in the $\sigma-\phi$ plane.  Periodic contributions can naturally arise in inflationary models with axions, such as in axion monodromy inflation (see e.g.~\cite{Baumann:2014nda} for review).
The tilt causes the scalar field to roll slowly along the rim of the ``hat'' and source inflation until it stops at the minimum. Since the dent in the middle of the ``hat'' sits at a deeper minimum for a sizable portion of the path than the slow-rolling field separated by a barrier, the field can tunnel to this vacuum\footnote{In de Sitter space, tunneling to a higher energy vacuum is also allowed, but the rate is suppressed~\cite{Lee:1987qc}.}. For a sufficiently small tilt, the bubble nucleation rate $\lambda \sim e^{-S_E}$ that depends on the Euclidian instanton bounce action for the vacuum tunneling  $S_E$
 is approximately constant and for specific model parameters can be computed from the bounce action using well known techniques~\cite{Coleman:1985ki, Kusenko:1995jv}. Considering thin-wall approximation and keeping the terms in the expansion of Eq.~\eqref{eq:model} potential up to order six, the action can be estimated\footnote{We thank the anonymous referee for suggestion.} as $S_E \sim 0.3 (a/c)^{13/2} c^2 f^3/g^3$.  Imposing requirements on the tunneling rate, the size of quantum  fluctuations as well as the duration of the inflationary period will introduce additional fine-tuning of the model parameters~\cite{Garriga:2015fdk,Deng:2017uwc,Deng:2018cxb}. As usual with models of inflation, some fine-tuning of the tilt of the Mexican hat is necessary to ensure the slow roll.  An independent set of parameters controls the tunneling rate, and these parameters determine the position of the mass function and the PBH abundance.

The tunneling rate becomes increasingly suppressed and effectively shuts off as the field rolls towards the portion of the tilted rim whose height is deeper than the minimum of the dent at the origin. Below, we take $M_{\rm pl} = 1$.
 
The resulting bubbles with the energy density $\rho_b=V(0,0)$ in their interior have a radius smaller than the inflationary Hubble length $H_i^{-1} = (8 \pi   \rho_i/3)^{-1/2}$ at the time of formation. The pressure $P = \rho_i - \rho_b$ on the wall causes the bubble to expand until $P$ changes sign as $\rho_i$ decreases below $\rho_b$.  They  undergo rapid expansion until the energy density inside the bubble exceeds the energy density in the exterior, which happens at some point before 
the end of inflation at time $t_i$. After that, the bubble contracts and collapses to a black hole. Interactions with the surrounding medium  can also affect the bubble wall momentum during the last stages of expansion.

While for the outside observer residing in the parent Friedmann-Lemaitre-Robertson-Walker  Universe the result of a bubble evolution is a black hole, the dynamics of the bubble interior depend on whether the bubble radius $R$ exceeds $H_b^{-1} = (8 \pi   \rho_b/3)^{-1/2}$ during expansion~\cite{Blau:1986cw}. If $R < H_b^{-1}$ at all times, the bubble is subcritical, and it will eventually collapse to a black hole
under the effects of vacuum pressure, wall tension and radiation pressure. At the end of inflation, when the Hubble radius is $t_i \sim H_i^{-1}$, the bubble radius is $R_i$. Prior to thermalization, the energy of the region excluded by the bubble contains inflaton energy of $E_i = (4 \pi/3) \rho_i R_i^3$. The mass of the resulting black hole is approximately the energy of the bubble~\cite{Deng:2017uwc}: 
\begin{equation} \label{eq:subcr}
M \simeq E_b \simeq \Big(\dfrac{4 \pi}{3} \rho_b + 4 \pi \sigma H_i\Big) R_i^3 = \kappa R_i^3~,
\end{equation}
where $\sigma$ is the bubble wall tension and the first and second term represent the bubble energy density and wall energy contributions, respectively. In the presence of plasma from the inflaton decay, the energy difference $(E_i - E_b)$ is transferred to the outgoing shock wave powered by the radiation reflected from the bubble wall.

If $R > H_b^{-1}$ during inflation, the bubble is supercritical. In this case, the interior can support inflation driven by $\rho_b$ within a de Sitter horizon of size $H_b^{-1}$. This region is connected through a wormhole to the exterior of the bubble~\cite{Maeda:1981gw,Kodama:1981gu,Sato:1981bf,Deng:2017uwc}. Eventually, the link is broken and a separate ``baby universe'' is formed, leading to a multiverse structure~\cite{Sato:1981gv} reminiscent of eternal inflation~\cite{Linde:2015edk}. From causality, the region affected by the Schwarzschild radius of the black hole resulting from the bubble collapse cannot exceed the Hubble radius of the parent Universe $t_h = a(t_h)R_i$, where $a$ is the scale factor. In radiation-dominated era $a = (t/t_i)^{1/2}$ and $t_h = H_i R_i^2$. Numerical simulations confirm that the resulting black hole mass saturates this bound~\cite{Deng:2017uwc}
\begin{equation} \label{eq:supcr}
M \sim \dfrac{4 \pi}{3} \rho  (t_h) H^{-3}(t_h) = H_i R_i^2~. 
\end{equation}
The subcritical relation, Eq.~\eqref{eq:subcr}, does not apply when $R_i \gg H_i/\kappa$ or $M \gg M_{\ast} \sim H_i^3/\kappa^3$.  

At the end of inflation, the bubble sizes have a broad distribution  depending on the formation time\footnote{Assuming bubbles nucleate with initial radius that is negligible compared to $H_i^{-1}$, the future bubble radius is approximately independent of initial radii distribution and it will not affect PBH mass-function. Hence, the bubble radius and the bubble number density $n(R_i)$ are fixed by the end of inflation time~$t_i$. After $t_i$, the bubble population is diluted by cosmic expansion~\cite{Deng:2017uwc}.}.   The number density of the bubbles with radius $\sim R_i$ is $n(R_i) = \lambda R_i^{-3}$, where $\lambda$ is the dimensionless bubble nucleation rate per Hubble volume per Hubble time. Here we assume that variation of $\lambda$ is small, and it is approximately constant for some some time during the slow-roll evolution of the inflaton.  Using Eqs.~\eqref{eq:subcr} 
and \eqref{eq:supcr} one can obtain the mass function of PBHs normalized to the DM density: 
\begin{equation}
    f(M) = \dfrac{M^2}{\rho_{\rm DM}} \dfrac{dn}{dM}~,
\end{equation}
where $\rho_{\rm DM}$ is the dark matter density, which scales as $\rho_{\rm DM} (t) \sim (B t^{3/2} M_{\rm eq}^{1/2})^{-1}$ during radiation era $t < t_{\rm eq}$ prior to matter-radiation equality, $B \sim 10$ is a numerical coefficient, and $M_{\rm eq} \sim 10^{17} M_{\odot}$ is the horizon mass at $t_{\rm eq}$. This results in a broad mass function~\cite{Deng:2017uwc}: 
\begin{equation} \label{eq:pbhspec}
f(M) \sim B \lambda M_{\rm eq}^{1/2} 
\begin{cases}
  M_{\ast}^{-1/2} & \text{for }M< M_{\ast}\\ M^{-1/2} & \text{for }M> M_{\ast}~.\\ 
\end{cases}
\end{equation}
The distribution $f(M)$ has an effective lower cutoff at $M_{\rm min} \sim \kappa H_i^{-3}$, when $R_{\rm min} < H_i^{-1}$. Thus, the total fraction of PBH in DM is
\begin{equation}
f_{\rm PBH} \sim B \lambda \Big(\dfrac{M_{\rm eq}}{M_{\ast}}\Big)^{1/2}\Big[\log\Big(\dfrac{M_{\ast}}{M_{\rm min}}\Big) + 1\Big]~. 
\end{equation}
At the lower end of the spectrum quantum fluctuations suppress black hole formation. The upper cut-off is very large and is set by $R_i < H_i^{-1} e^{N}$, where $N \sim 60$ is the number of $e$-folds of inflation during which the bubble nucleation takes place. We note that while above $\lambda$ was approximated by a constant, in models with a potential of the form~\eqref{eq:model}, the tunneling rate slowly varies, and, therefore, the cutoff in $f(M)$ is not a step function, but a smooth function corresponding to the exponential suppression of tunneling $\sim e^{-S_E}$.

\begin{figure}[tb]
  \includegraphics[width=0.9\linewidth]{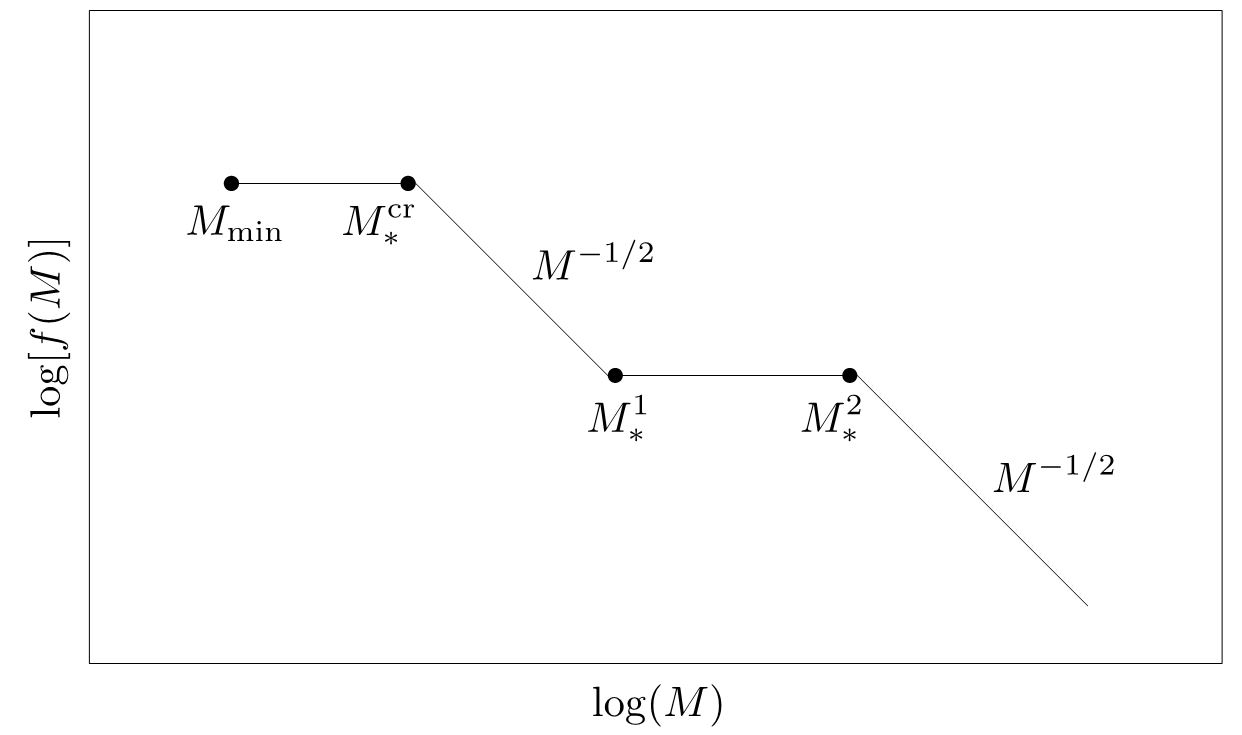}
\caption{Schematic illustration of the PBH mass spectrum from vacuum bubbles with an intermediate matter-dominated era.}
\label{fig:genspec}
\end{figure}

While Refs.~\cite{Deng:2017uwc,Deng:2016vzb} 
focused on PBH formation in radiation-dominated era, it is possible and indeed likely that inflation is followed by an era of coherent oscillations of the inflaton, during which the expansion rate is the same as in a matter-dominated phase~\cite{Garriga:2015fdk}.  An intermediate matter dominated era can also be caused by moduli or spectator fields, or by a fragmentation of a scalar field into solitonic lumps~\cite{Kusenko:1997si,Cotner:2016cvr,Cotner:2017tir,Cotner:2018vug,Cotner:2019ykd}. While for subcritical bubbles the results of Eq.~\eqref{eq:pbhspec} are not 
affected, PBHs from supercritical bubbles formed during this era exhibit a different mass scaling than that in Eq.~\eqref{eq:supcr}.  During the matter-dominated era $a = (t/t_i)^{2/3}$ and $t_h = H_i^2 R_i^3$, and, therefore, the black hole mass from supercritical bubbles scales as $M \sim R_i^3$, instead of $M \sim R_i^2$. We can now generalize the PBH mass function of Eq.~\eqref{eq:pbhspec} to 
\begin{equation} \label{eq:pbhspec2}
f(M) \sim B \lambda M_{\rm eq}^{1/2} 
\begin{cases}
  (M_{\ast}^{\rm cr})^{-1/2} & \text{for }M< M_{\ast}^{\rm cr}\\ M^{-1/2} & \text{for }M_{\ast}^1 > M> M_{\ast}^{\rm cr} \\
  (M_{\ast}^1)^{-1/2} & \text{for } M_{\ast}^2 > M> M_{\ast}^1  \\
    M^{-1/2} & \text{for } M> M_{\ast}^2~,  \\ 
\end{cases}
\end{equation}
where $M_{\ast}^{\rm cr}$ denotes transition between super and subcritical bubbles as before, while $M_{\ast}^1$ and $M_{\ast}^2$ denote the beginning and the end of the intermediate matter-dominated  phase.  We display the resulting PBH mass spectrum in Fig.~\ref{fig:genspec}. The above can be readily extended to include an arbitrary number of such radiation-matter transitions. Since the values of $M_{\rm min}, M_{\ast}^{\rm cr}, M_{\ast}^1, M_{\ast}^2$ and $\lambda$ depend on the particle model, we take them as free parameters.

The range of PBH masses is limited from above by the temperature of the latest reheating at the end of the last intermediate matter-dominated phase.  One constraint is that the reheat temperature may not be lower than a few MeV for Big Bang nucleosynthesis to take place.  Another potential constraint is imposed by baryon asymmetry of the Universe. 
In the scenario with a single radiation-dominated era, as in Eq.~\eqref{eq:pbhspec}, baryogenesis can take place at a high temperature, as typically considered. On the other hand, when there is an intermediate matter-dominated era as in Eq.~\eqref{eq:pbhspec2}, the large PBH masses imply a low reheat temperature.  To produce PBH masses of the order of the solar mass or larger, one must assume that the reheat temperature after the intermediate matter-dominated phase is as low as  $T_r \sim$ GeV.  Any baryon asymmetry produced before the intermediate matter-dominated era will be diluted by a large factor $\gtrsim 10^8$. A low-scale baryogenesis required in this case can occur via scalar curvaton field and Affleck-Dine mechanism~\cite{Affleck:1984fy,Dine:2003ax}, or late-decaying moduli~(e.g.~\cite{Allahverdi:2010im,Kitano:2008tk,Chen:2018uzu}).

If the PBHs form during a radiation-dominated era, the expanding bubbles generate shock waves and sound waves.  Their effects are not  entirely dissipated by Silk damping, and they can leave an imprint on the cosmic microwave background (CMB) through the $\mu$-distortions, imposing a restriction on normalization of the $\propto M^{-1/2}$ tail of the PBH spectrum related to the bubble nucleation rate $\lambda \lesssim 10^{-15}$~\cite{Deng:2017uwc,Deng:2018cxb}. However, this constraint relies on the assumption that expanding bubble walls interact  with radiation and plasma.  If the bubble expansion takes place during a matter-dominated phase, the constraint does not apply.  

The broad and multistep PBH spectrum shape of Eq.~\eqref{eq:pbhspec2} allows us to naturally explain an extensive range of phenomena simultaneously within a single model, which cannot be accomplished with the spectrum of Eq.~\eqref{eq:pbhspec}. PBHs can account for all DM if  $M_{\rm min} = M_{\ast}^{\rm cr}$ lies in the open parameter window of $\sim 10^{-16} - 10^{-8} M_{\odot}$.  Observed LIGO events can be caused by PBHs if  $f(M \sim 30 M_{\odot}) \sim 10^{-3}$~\cite{Sasaki:2016jop}, which we identify with $f(M_{\ast}^2)$. For PBHs to seed supermassive black holes one needs a black hole of $M \gtrsim 10^3 M_{\odot}$  in each galactic halo, corresponding to a density of $n_{M} \sim 0.1 $ Mpc$^{-3}$, which is possible if $\lambda \gtrsim 10^{-17}$ for $M \gtrsim 10^{3} M_{\odot}$. 

\begin{figure*}[tb]
\centering
  \centering
  \includegraphics[width=1.0\linewidth]{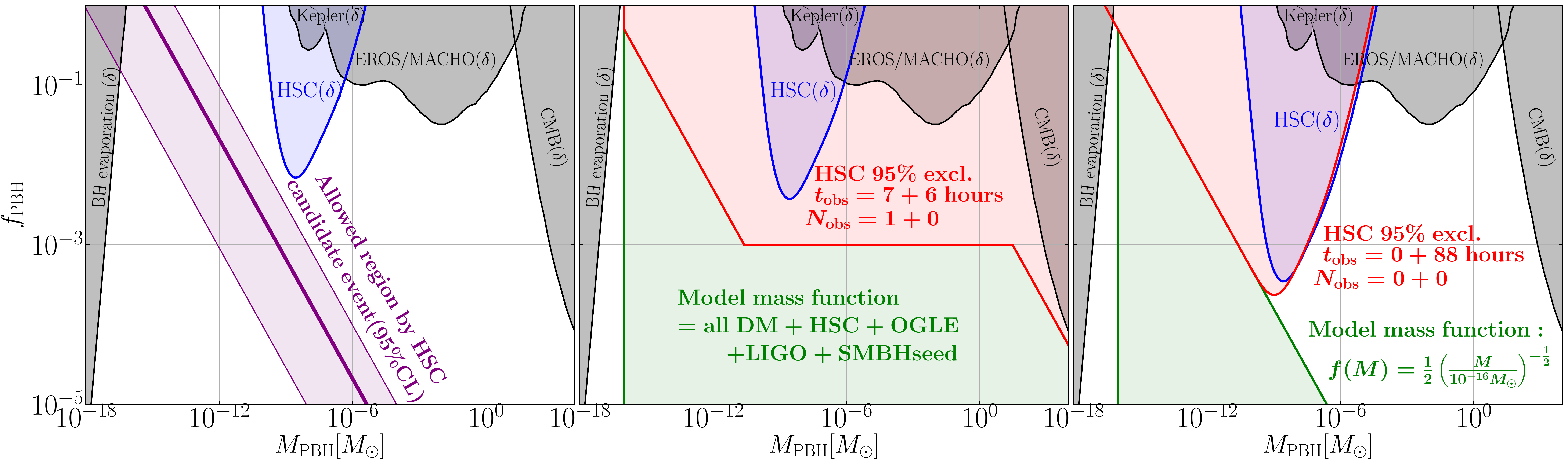}
\caption{[Left] Allowed normalization range for a PBH mass function $\propto M^{-1/2}$ to be consistent with the HSC candidate event reported after 7 hours of observations~\cite{Niikura:2017zjd}. The HSC constraint  (shaded blue) takes into account the updated finite-source size effects~\cite{Smyth:2019whb,Sugiyama:2019dgt}. Thick purple line represents the best fit and the band corresponds to a 95\% confidence level (CL) interval. For each line in the allowed range, the mass function can be made consistent with $f_{\rm{PBH}}=1$ by introducing a low-mass  cutoff in the $10^{-15}-10^{-10}M_\odot$ range.
[Middle] Green line shows a model mass function, in which PBHs can account for HSC and OGLE microlensing observation, LIGO observations~\cite{Sasaki:2016jop} (this region will be further tested with stochastic gravity-wave background~\cite{Wang:2016ana,Nakamura:1997sm}) and seeds of supermassive black holes. After 6 hours of additional observation by HSC, we can exclude $f_{\rm PBH}=1$ normalization (red region), assuming null detection in future observation. 
[Right] Green line shows the most pessimistic scenario, corresponding to the lowest possible normalization and $M_{\rm min} \sim 10^{-16}M_\odot$, for which PBH can still account for all of the DM. The red region is the exclusion region of normalization $f_{\rm PBH}$ at a given $M_{\rm min}$, assuming 88 hours of new observation and null detection. The blue region in each panel is the constraint assuming monotonic mass function and corresponding observation time (7, 7$+$6, and 88 hours from left to right).
Constraints from  extragalactic $\gamma$-rays from BH evaporation~\cite{Carr:2009jm} (additional constraints in this region due to positron production from BH evaporation have been also recently suggested \cite{Dasgupta:2019cae,Laha:2019ssq,DeRocco:2019fjq}), microlensing Kepler data~\cite{Griest:2013aaa}, MACHO, EROS and OGLE microlensing~\cite{Tisserand:2006zx}, and the accretion effects on the CMB observables~\cite{Ali-Haimoud:2016mbv} (see also~Ref.~\cite{Poulin:2017bwe}) are also displayed. The label of $(\delta)$ denotes that the constraint is derived assuming monotonic mass function.}
\label{fig:finalres}
\end{figure*}

Furthermore, HSC microselensing observations of the Andromeda galaxy (M31)~\cite{Niikura:2017zjd} reported a candidate event consistent with PBHs at $f(M \sim 10^{-9} M_{\odot}) \sim 10^{-2}$. It has been suggested in Ref.~\cite{Deng:2018cxb} that a broad PBH spectrum from vacuum bubbles of Eq.~\eqref{eq:pbhspec} can accommodate this as well as DM. 

Given the exciting possibility that all of these phenomena might be explained by PBHs produced from bubble nucleation, it is important to explore the discovery range of the HSC.  We study the HSC reach numerically, and we find that upcoming observations of the HSC will allow to fully test vacuum bubbles as the primary source of PBH DM. Furthermore, the HSC will be able to probe the intriguing scenario represented by Eq.~\eqref{eq:pbhspec2} that can simultaneously explain LIGO events and SMBH seeds,  while PBHs from vacuum bubbles constitute all of the DM.
 
We employ results from HSC Monte Carlo simulations as well as their analysis tools, outlined in Ref.~\cite{Niikura:2017zjd}, to perform a fit of the PBH mass-spectrum to the expected number of observed microlensing events
\begin{widetext}
\begin{equation}
N_{\rm exp} \Big(\dfrac{\Omega_{\rm PBH}}{\Omega_{\rm DM}}\Big) = \dfrac{\Omega_{\rm PBH}}{\Omega_{\rm DM}} \int\! dM \int_0^{t_{\rm obs}}\! \dfrac{\mathrm{d}t_{\rm FWHM}}{t_{\rm FWHM}} \int\! \mathrm{d}m_r \dfrac{\mathrm{d}N_{\rm event}}{\mathrm{d} \log(t_{\rm FWHM})} \dfrac{\mathrm{d}N_s}{\mathrm{d}m_r} \epsilon(t_{\rm FWHM}, m_r) \dfrac{f(M)}{M}~,
\end{equation}
\end{widetext}
where $(\Omega_{\rm PBH}/\Omega_{\rm DM}) = f_{\rm PBH}$ is the 
{mass} fraction of DM in the form of PBHs, $\mathrm{d}N_{\rm event}/\mathrm{d} \log(t_{\rm FWHM})$ is the expected differential number of PBH microlensing events per logarithmic interval of the fullwidth-at-half-maximum (FWHM) microlensing timescale $t_{\rm FWHM}$ for a single star in M31, 
$\mathrm{d}N_s/\mathrm{d}m_r$ is the luminosity function of source stars in the photometric $r$-band magnitude range $[m_r, r + \mathrm{d}m_r]$, $\epsilon(t_{\rm FWHM}, m_r)$ is the detection efficiency quantifying the probability that a microlensing event for a star
with magnitude $m_r$ and the light curve timescale $t_{\rm FWHM}$ is detected by HSC event selection procedures and the PBH mass spectrum $f(M)$ is normalized to satisfy $\int_0^{\infty}dM f(M)/M = 1$.

We first analyze compatibility of the broad PBH spectrum described by $f(M)$ with detection of a single candidate event reported by the HSC after 7 hours of observations.  The mass function must be consistent with one event corresponding to  the candidate PBH mass, while no events are observed at other masses.  The $f(M) \propto M^{-1/2}$ mass function passes this test in the range of normalizations shown in Fig.~\ref{fig:finalres}, leftmost panel. (We note in passing that, since  the PBH spectrum is not monochromatic, the lines in the allowed range do not reach the differential HSC exclusion region, but pass notably lower.)  Furthermore, for each line  in the allowed range, one can obtain $f_{\rm{PBH}}=1 $ by introducing a low-mass cutoff in the allowed range $(10^{-15}-10^{-10})M_\odot$. 

To explore the HSC reach to probe PBH DM from vacuum bubbles, we estimate the required time for upcoming HSC observations to start seeing events. The results are displayed in Fig.~\ref{fig:finalres}. For the general model with the choice of parameters that can simultaneously explain all of the DM, LIGO events and SMBHs (middle panel), we find that HSC already started to probe this scenario, and new detections can be expected with only 2.7 hours of observations. Based on Poisson statistics, a single HSC candidate event found after 7 hours of observation is still compatible with this scenario at $\sim 19\%$ C.L.. 
The scenario of $f_{\rm PBH}=1$ with fixed shape of $f(M)$ can be excluded with additional 6-hours observation at a 2-$\sigma$ level (95\% C.L.), combining existing 7 hours of observation and assuming null detection in future observation. The red shaded region is the exclusion region after 13 hours of observation in total.

The HSC reach for the most pessimistic realization of the vacuum bubble PBH DM scenario, corresponding to normalization with the lowest possible nucleation rate $\lambda$, is also impressive (rightmost panel). 
We find that 88 hours of future observation can exclude the scenario at a 2-$\sigma$ level, assuming null detection. The red shaded region is exclusion region of $f_{\rm PBH}$ at a given cutoff scale, $M_{\rm min}$. For $M_{\rm min} \lesssim 10^{-11}M_\odot$, the constraining power saturates because every $M_{\rm min}\lesssim 10^{-11}M_\odot$ gives same number of microlensing events.

Another promising microlensing observatory will be the Rubin Observatory Legacy Survey of Space and Time (LSST)\footnote{\url{https://www.lsst.org}}, which is expected to start its full science operation in 2022. If LSST carries out a microlensing survey towards the Galactic Center that is accessible from the LSST site in Chile, it would easily test the PBH scenario, thanks to its large mirror aperture, wide field of view, higher detector sensitivity, and the expected superb image quality that allow for a simultaneous monitoring observation of many stars at one time, just as the Subaru HSC does for M31 (see also~Ref.~\cite{Niikura:2019kqi} for a similar discussion). An optimal cadence strategy needs to be explored in order to maximize science outputs of microlensing observations to constrain the abundance of not only PBHs, but of astrophysical compact objects (neutron stars and black holes) as well~\cite{Abrams:2020jvs}.

In conclusion, we have presented a general scenario of PBH formation from vacuum bubbles and discussed its intriguing realization that can naturally account for all of the dark matter, observed LIGO events as well as seeds of supermassive black holes within a single model. While PBH DM with masses in the open parameter space window is difficult to test, the tail of the distribution extending to larger masses makes it possible to probe this exciting possibility with the HSC. We used detailed numerical tools to show that upcoming HSC observations, as well as the future observations with LSST, will allow us to definitively test the general PBH formation scenario from vacuum bubbles as the primary source of DM. 

The work of A.K., V.T. and E.V. was supported by the U.S. Department of Energy (DOE) Grant No. DE-SC0009937.  A.K. was also supported by the World Premier International Research Center Initiative (WPI), MEXT, Japan. This research was also supported in part by the National Science Foundation under Grant No. NSF PHY-1748958. This work was supported in part by JSPS KAKENHI Grant Numbers JP15H03654, JP15H05887, JP15H05888, JP15H05893, JP15H05896, JP15K21733, and JP19H00677.

 
\bibliography{bibliography}

 \newcommand{\noop}[1]{}
\begin{thebibliography}{80}%
\makeatletter
\providecommand \@ifxundefined [1]{%
 \@ifx{#1\undefined}
}%
\providecommand \@ifnum [1]{%
 \ifnum #1\expandafter \@firstoftwo
 \else \expandafter \@secondoftwo
 \fi
}%
\providecommand \@ifx [1]{%
 \ifx #1\expandafter \@firstoftwo
 \else \expandafter \@secondoftwo
 \fi
}%
\providecommand \natexlab [1]{#1}%
\providecommand \enquote  [1]{``#1''}%
\providecommand \bibnamefont  [1]{#1}%
\providecommand \bibfnamefont [1]{#1}%
\providecommand \citenamefont [1]{#1}%
\providecommand \href@noop [0]{\@secondoftwo}%
\providecommand \href [0]{\begingroup \@sanitize@url \@href}%
\providecommand \@href[1]{\@@startlink{#1}\@@href}%
\providecommand \@@href[1]{\endgroup#1\@@endlink}%
\providecommand \@sanitize@url [0]{\catcode `\\12\catcode `\$12\catcode
  `\&12\catcode `\#12\catcode `\^12\catcode `\_12\catcode `\%12\relax}%
\providecommand \@@startlink[1]{}%
\providecommand \@@endlink[0]{}%
\providecommand \url  [0]{\begingroup\@sanitize@url \@url }%
\providecommand \@url [1]{\endgroup\@href {#1}{\urlprefix }}%
\providecommand \urlprefix  [0]{URL }%
\providecommand \Eprint [0]{\href }%
\providecommand \doibase [0]{http://dx.doi.org/}%
\providecommand \selectlanguage [0]{\@gobble}%
\providecommand \bibinfo  [0]{\@secondoftwo}%
\providecommand \bibfield  [0]{\@secondoftwo}%
\providecommand \translation [1]{[#1]}%
\providecommand \BibitemOpen [0]{}%
\providecommand \bibitemStop [0]{}%
\providecommand \bibitemNoStop [0]{.\EOS\space}%
\providecommand \EOS [0]{\spacefactor3000\relax}%
\providecommand \BibitemShut  [1]{\csname bibitem#1\endcsname}%
\let\auto@bib@innerbib\@empty
\bibitem [{\citenamefont {{Zel'dovich}}\ and\ \citenamefont
  {{Novikov}}(1967)}]{Zeldovich:1967}%
  \BibitemOpen
  \bibfield  {author} {\bibinfo {author} {\bibfnamefont {Y.~B.}\ \bibnamefont
  {{Zel'dovich}}}\ and\ \bibinfo {author} {\bibfnamefont {I.~D.}\ \bibnamefont
  {{Novikov}}},\ }\href@noop {} {\bibfield  {journal} {\bibinfo  {journal}
  {Sov. Astron.}\ }\textbf {\bibinfo {volume} {10}},\ \bibinfo {pages} {602}
  (\bibinfo {year} {1967})}\BibitemShut {NoStop}%
\bibitem [{\citenamefont {Hawking}(1971)}]{Hawking:1971ei}%
  \BibitemOpen
  \bibfield  {author} {\bibinfo {author} {\bibfnamefont {S.}~\bibnamefont
  {Hawking}},\ }\href@noop {} {\bibfield  {journal} {\bibinfo  {journal} {Mon.
  Not. Roy. Astron. Soc.}\ }\textbf {\bibinfo {volume} {152}},\ \bibinfo
  {pages} {75} (\bibinfo {year} {1971})}\BibitemShut {NoStop}%
\bibitem [{\citenamefont {Carr}\ and\ \citenamefont
  {Hawking}(1974)}]{Carr:1974nx}%
  \BibitemOpen
  \bibfield  {author} {\bibinfo {author} {\bibfnamefont {B.~J.}\ \bibnamefont
  {Carr}}\ and\ \bibinfo {author} {\bibfnamefont {S.~W.}\ \bibnamefont
  {Hawking}},\ }\href@noop {} {\bibfield  {journal} {\bibinfo  {journal} {Mon.
  Not. Roy. Astron. Soc.}\ }\textbf {\bibinfo {volume} {168}},\ \bibinfo
  {pages} {399} (\bibinfo {year} {1974})}\BibitemShut {NoStop}%
\bibitem [{\citenamefont {Garcia-Bellido}\ \emph {et~al.}(1996)\citenamefont
  {Garcia-Bellido}, \citenamefont {Linde},\ and\ \citenamefont
  {Wands}}]{GarciaBellido:1996qt}%
  \BibitemOpen
  \bibfield  {author} {\bibinfo {author} {\bibfnamefont {J.}~\bibnamefont
  {Garcia-Bellido}}, \bibinfo {author} {\bibfnamefont {A.~D.}\ \bibnamefont
  {Linde}}, \ and\ \bibinfo {author} {\bibfnamefont {D.}~\bibnamefont
  {Wands}},\ }\href {\doibase 10.1103/PhysRevD.54.6040} {\bibfield  {journal}
  {\bibinfo  {journal} {Phys. Rev.}\ }\textbf {\bibinfo {volume} {D54}},\
  \bibinfo {pages} {6040} (\bibinfo {year} {1996})},\ \Eprint
  {http://arxiv.org/abs/astro-ph/9605094} {arXiv:astro-ph/9605094 [astro-ph]}
  \BibitemShut {NoStop}%
\bibitem [{\citenamefont {Khlopov}(2010)}]{Khlopov:2008qy}%
  \BibitemOpen
  \bibfield  {author} {\bibinfo {author} {\bibfnamefont {M.~{\relax Yu}.}\
  \bibnamefont {Khlopov}},\ }\href {\doibase 10.1088/1674-4527/10/6/001}
  {\bibfield  {journal} {\bibinfo  {journal} {Res. Astron. Astrophys.}\
  }\textbf {\bibinfo {volume} {10}},\ \bibinfo {pages} {495} (\bibinfo {year}
  {2010})},\ \Eprint {http://arxiv.org/abs/0801.0116} {arXiv:0801.0116
  [astro-ph]} \BibitemShut {NoStop}%
\bibitem [{\citenamefont {Frampton}\ \emph {et~al.}(2010)\citenamefont
  {Frampton}, \citenamefont {Kawasaki}, \citenamefont {Takahashi},\ and\
  \citenamefont {Yanagida}}]{Frampton:2010sw}%
  \BibitemOpen
  \bibfield  {author} {\bibinfo {author} {\bibfnamefont {P.~H.}\ \bibnamefont
  {Frampton}}, \bibinfo {author} {\bibfnamefont {M.}~\bibnamefont {Kawasaki}},
  \bibinfo {author} {\bibfnamefont {F.}~\bibnamefont {Takahashi}}, \ and\
  \bibinfo {author} {\bibfnamefont {T.~T.}\ \bibnamefont {Yanagida}},\ }\href
  {\doibase 10.1088/1475-7516/2010/04/023} {\bibfield  {journal} {\bibinfo
  {journal} {JCAP}\ }\textbf {\bibinfo {volume} {1004}},\ \bibinfo {pages}
  {023} (\bibinfo {year} {2010})},\ \Eprint {http://arxiv.org/abs/1001.2308}
  {arXiv:1001.2308 [hep-ph]} \BibitemShut {NoStop}%
\bibitem [{\citenamefont {Kawasaki}\ \emph {et~al.}(2016)\citenamefont
  {Kawasaki}, \citenamefont {Kusenko}, \citenamefont {Tada},\ and\
  \citenamefont {Yanagida}}]{Kawasaki:2016pql}%
  \BibitemOpen
  \bibfield  {author} {\bibinfo {author} {\bibfnamefont {M.}~\bibnamefont
  {Kawasaki}}, \bibinfo {author} {\bibfnamefont {A.}~\bibnamefont {Kusenko}},
  \bibinfo {author} {\bibfnamefont {Y.}~\bibnamefont {Tada}}, \ and\ \bibinfo
  {author} {\bibfnamefont {T.~T.}\ \bibnamefont {Yanagida}},\ }\href {\doibase
  10.1103/PhysRevD.94.083523} {\bibfield  {journal} {\bibinfo  {journal} {Phys.
  Rev.}\ }\textbf {\bibinfo {volume} {D94}},\ \bibinfo {pages} {083523}
  (\bibinfo {year} {2016})},\ \Eprint {http://arxiv.org/abs/1606.07631}
  {arXiv:1606.07631 [astro-ph.CO]} \BibitemShut {NoStop}%
\bibitem [{\citenamefont {Cotner}\ and\ \citenamefont
  {Kusenko}(2017{\natexlab{a}})}]{Cotner:2016cvr}%
  \BibitemOpen
  \bibfield  {author} {\bibinfo {author} {\bibfnamefont {E.}~\bibnamefont
  {Cotner}}\ and\ \bibinfo {author} {\bibfnamefont {A.}~\bibnamefont
  {Kusenko}},\ }\href {\doibase 10.1103/PhysRevLett.119.031103} {\bibfield
  {journal} {\bibinfo  {journal} {Phys. Rev. Lett.}\ }\textbf {\bibinfo
  {volume} {119}},\ \bibinfo {pages} {031103} (\bibinfo {year}
  {2017}{\natexlab{a}})},\ \Eprint {http://arxiv.org/abs/1612.02529}
  {arXiv:1612.02529 [astro-ph.CO]} \BibitemShut {NoStop}%
\bibitem [{\citenamefont {Cotner}\ and\ \citenamefont
  {Kusenko}(2017{\natexlab{b}})}]{Cotner:2017tir}%
  \BibitemOpen
  \bibfield  {author} {\bibinfo {author} {\bibfnamefont {E.}~\bibnamefont
  {Cotner}}\ and\ \bibinfo {author} {\bibfnamefont {A.}~\bibnamefont
  {Kusenko}},\ }\href {\doibase 10.1103/PhysRevD.96.103002} {\bibfield
  {journal} {\bibinfo  {journal} {Phys. Rev.}\ }\textbf {\bibinfo {volume}
  {D96}},\ \bibinfo {pages} {103002} (\bibinfo {year} {2017}{\natexlab{b}})},\
  \Eprint {http://arxiv.org/abs/1706.09003} {arXiv:1706.09003 [astro-ph.CO]}
  \BibitemShut {NoStop}%
\bibitem [{\citenamefont {Carr}\ \emph {et~al.}(2016)\citenamefont {Carr},
  \citenamefont {Kuhnel},\ and\ \citenamefont {Sandstad}}]{Carr:2016drx}%
  \BibitemOpen
  \bibfield  {author} {\bibinfo {author} {\bibfnamefont {B.}~\bibnamefont
  {Carr}}, \bibinfo {author} {\bibfnamefont {F.}~\bibnamefont {Kuhnel}}, \ and\
  \bibinfo {author} {\bibfnamefont {M.}~\bibnamefont {Sandstad}},\ }\href
  {\doibase 10.1103/PhysRevD.94.083504} {\bibfield  {journal} {\bibinfo
  {journal} {Phys. Rev.}\ }\textbf {\bibinfo {volume} {D94}},\ \bibinfo {pages}
  {083504} (\bibinfo {year} {2016})},\ \Eprint
  {http://arxiv.org/abs/1607.06077} {arXiv:1607.06077 [astro-ph.CO]}
  \BibitemShut {NoStop}%
\bibitem [{\citenamefont {Inomata}\ \emph
  {et~al.}(2017{\natexlab{a}})\citenamefont {Inomata}, \citenamefont
  {Kawasaki}, \citenamefont {Mukaida}, \citenamefont {Tada},\ and\
  \citenamefont {Yanagida}}]{Inomata:2016rbd}%
  \BibitemOpen
  \bibfield  {author} {\bibinfo {author} {\bibfnamefont {K.}~\bibnamefont
  {Inomata}}, \bibinfo {author} {\bibfnamefont {M.}~\bibnamefont {Kawasaki}},
  \bibinfo {author} {\bibfnamefont {K.}~\bibnamefont {Mukaida}}, \bibinfo
  {author} {\bibfnamefont {Y.}~\bibnamefont {Tada}}, \ and\ \bibinfo {author}
  {\bibfnamefont {T.~T.}\ \bibnamefont {Yanagida}},\ }\href {\doibase
  10.1103/PhysRevD.95.123510} {\bibfield  {journal} {\bibinfo  {journal} {Phys.
  Rev. D}\ }\textbf {\bibinfo {volume} {95}},\ \bibinfo {pages} {123510}
  (\bibinfo {year} {2017}{\natexlab{a}})},\ \Eprint
  {http://arxiv.org/abs/1611.06130} {arXiv:1611.06130 [astro-ph.CO]}
  \BibitemShut {NoStop}%
\bibitem [{\citenamefont {Pi}\ \emph {et~al.}(2018)\citenamefont {Pi},
  \citenamefont {Zhang}, \citenamefont {Huang},\ and\ \citenamefont
  {Sasaki}}]{Pi:2017gih}%
  \BibitemOpen
  \bibfield  {author} {\bibinfo {author} {\bibfnamefont {S.}~\bibnamefont
  {Pi}}, \bibinfo {author} {\bibfnamefont {Y.-l.}\ \bibnamefont {Zhang}},
  \bibinfo {author} {\bibfnamefont {Q.-G.}\ \bibnamefont {Huang}}, \ and\
  \bibinfo {author} {\bibfnamefont {M.}~\bibnamefont {Sasaki}},\ }\href
  {\doibase 10.1088/1475-7516/2018/05/042} {\bibfield  {journal} {\bibinfo
  {journal} {JCAP}\ }\textbf {\bibinfo {volume} {1805}},\ \bibinfo {pages}
  {042} (\bibinfo {year} {2018})},\ \Eprint {http://arxiv.org/abs/1712.09896}
  {arXiv:1712.09896 [astro-ph.CO]} \BibitemShut {NoStop}%
\bibitem [{\citenamefont {Inomata}\ \emph
  {et~al.}(2017{\natexlab{b}})\citenamefont {Inomata}, \citenamefont
  {Kawasaki}, \citenamefont {Mukaida}, \citenamefont {Tada},\ and\
  \citenamefont {Yanagida}}]{Inomata:2017okj}%
  \BibitemOpen
  \bibfield  {author} {\bibinfo {author} {\bibfnamefont {K.}~\bibnamefont
  {Inomata}}, \bibinfo {author} {\bibfnamefont {M.}~\bibnamefont {Kawasaki}},
  \bibinfo {author} {\bibfnamefont {K.}~\bibnamefont {Mukaida}}, \bibinfo
  {author} {\bibfnamefont {Y.}~\bibnamefont {Tada}}, \ and\ \bibinfo {author}
  {\bibfnamefont {T.~T.}\ \bibnamefont {Yanagida}},\ }\href {\doibase
  10.1103/PhysRevD.96.043504} {\bibfield  {journal} {\bibinfo  {journal} {Phys.
  Rev. D}\ }\textbf {\bibinfo {volume} {96}},\ \bibinfo {pages} {043504}
  (\bibinfo {year} {2017}{\natexlab{b}})},\ \Eprint
  {http://arxiv.org/abs/1701.02544} {arXiv:1701.02544 [astro-ph.CO]}
  \BibitemShut {NoStop}%
\bibitem [{\citenamefont {Garcia-Bellido}\ \emph {et~al.}(2017)\citenamefont
  {Garcia-Bellido}, \citenamefont {Peloso},\ and\ \citenamefont
  {Unal}}]{Garcia-Bellido:2017aan}%
  \BibitemOpen
  \bibfield  {author} {\bibinfo {author} {\bibfnamefont {J.}~\bibnamefont
  {Garcia-Bellido}}, \bibinfo {author} {\bibfnamefont {M.}~\bibnamefont
  {Peloso}}, \ and\ \bibinfo {author} {\bibfnamefont {C.}~\bibnamefont
  {Unal}},\ }\href {\doibase 10.1088/1475-7516/2017/09/013} {\bibfield
  {journal} {\bibinfo  {journal} {JCAP}\ }\textbf {\bibinfo {volume} {1709}},\
  \bibinfo {pages} {013} (\bibinfo {year} {2017})},\ \Eprint
  {http://arxiv.org/abs/1707.02441} {arXiv:1707.02441 [astro-ph.CO]}
  \BibitemShut {NoStop}%
\bibitem [{\citenamefont {Inoue}\ and\ \citenamefont
  {Kusenko}(2017)}]{Inoue:2017csr}%
  \BibitemOpen
  \bibfield  {author} {\bibinfo {author} {\bibfnamefont {Y.}~\bibnamefont
  {Inoue}}\ and\ \bibinfo {author} {\bibfnamefont {A.}~\bibnamefont
  {Kusenko}},\ }\href {\doibase 10.1088/1475-7516/2017/10/034} {\bibfield
  {journal} {\bibinfo  {journal} {JCAP}\ }\textbf {\bibinfo {volume} {1710}},\
  \bibinfo {pages} {034} (\bibinfo {year} {2017})},\ \Eprint
  {http://arxiv.org/abs/1705.00791} {arXiv:1705.00791 [astro-ph.CO]}
  \BibitemShut {NoStop}%
\bibitem [{\citenamefont {Georg}\ and\ \citenamefont
  {Watson}(2017)}]{Georg:2017mqk}%
  \BibitemOpen
  \bibfield  {author} {\bibinfo {author} {\bibfnamefont {J.}~\bibnamefont
  {Georg}}\ and\ \bibinfo {author} {\bibfnamefont {S.}~\bibnamefont {Watson}},\
  }\href {\doibase 10.1007/JHEP09(2017)138} {\bibfield  {journal} {\bibinfo
  {journal} {JHEP}\ }\textbf {\bibinfo {volume} {09}},\ \bibinfo {pages} {138}
  (\bibinfo {year} {2017})},\ \Eprint {http://arxiv.org/abs/1703.04825}
  {arXiv:1703.04825 [astro-ph.CO]} \BibitemShut {NoStop}%
\bibitem [{\citenamefont {Inomata}\ \emph {et~al.}(2018)\citenamefont
  {Inomata}, \citenamefont {Kawasaki}, \citenamefont {Mukaida},\ and\
  \citenamefont {Yanagida}}]{Inomata:2017vxo}%
  \BibitemOpen
  \bibfield  {author} {\bibinfo {author} {\bibfnamefont {K.}~\bibnamefont
  {Inomata}}, \bibinfo {author} {\bibfnamefont {M.}~\bibnamefont {Kawasaki}},
  \bibinfo {author} {\bibfnamefont {K.}~\bibnamefont {Mukaida}}, \ and\
  \bibinfo {author} {\bibfnamefont {T.~T.}\ \bibnamefont {Yanagida}},\ }\href
  {\doibase 10.1103/PhysRevD.97.043514} {\bibfield  {journal} {\bibinfo
  {journal} {Phys. Rev. D}\ }\textbf {\bibinfo {volume} {97}},\ \bibinfo
  {pages} {043514} (\bibinfo {year} {2018})},\ \Eprint
  {http://arxiv.org/abs/1711.06129} {arXiv:1711.06129 [astro-ph.CO]}
  \BibitemShut {NoStop}%
\bibitem [{\citenamefont {Kocsis}\ \emph {et~al.}(2018)\citenamefont {Kocsis},
  \citenamefont {Suyama}, \citenamefont {Tanaka},\ and\ \citenamefont
  {Yokoyama}}]{Kocsis:2017yty}%
  \BibitemOpen
  \bibfield  {author} {\bibinfo {author} {\bibfnamefont {B.}~\bibnamefont
  {Kocsis}}, \bibinfo {author} {\bibfnamefont {T.}~\bibnamefont {Suyama}},
  \bibinfo {author} {\bibfnamefont {T.}~\bibnamefont {Tanaka}}, \ and\ \bibinfo
  {author} {\bibfnamefont {S.}~\bibnamefont {Yokoyama}},\ }\href {\doibase
  10.3847/1538-4357/aaa7f4} {\bibfield  {journal} {\bibinfo  {journal}
  {Astrophys. J.}\ }\textbf {\bibinfo {volume} {854}},\ \bibinfo {pages} {41}
  (\bibinfo {year} {2018})},\ \Eprint {http://arxiv.org/abs/1709.09007}
  {arXiv:1709.09007 [astro-ph.CO]} \BibitemShut {NoStop}%
\bibitem [{\citenamefont {Ando}\ \emph {et~al.}(2018)\citenamefont {Ando},
  \citenamefont {Inomata}, \citenamefont {Kawasaki}, \citenamefont {Mukaida},\
  and\ \citenamefont {Yanagida}}]{Ando:2017veq}%
  \BibitemOpen
  \bibfield  {author} {\bibinfo {author} {\bibfnamefont {K.}~\bibnamefont
  {Ando}}, \bibinfo {author} {\bibfnamefont {K.}~\bibnamefont {Inomata}},
  \bibinfo {author} {\bibfnamefont {M.}~\bibnamefont {Kawasaki}}, \bibinfo
  {author} {\bibfnamefont {K.}~\bibnamefont {Mukaida}}, \ and\ \bibinfo
  {author} {\bibfnamefont {T.~T.}\ \bibnamefont {Yanagida}},\ }\href {\doibase
  10.1103/PhysRevD.97.123512} {\bibfield  {journal} {\bibinfo  {journal} {Phys.
  Rev. D}\ }\textbf {\bibinfo {volume} {97}},\ \bibinfo {pages} {123512}
  (\bibinfo {year} {2018})},\ \Eprint {http://arxiv.org/abs/1711.08956}
  {arXiv:1711.08956 [astro-ph.CO]} \BibitemShut {NoStop}%
\bibitem [{\citenamefont {Cotner}\ \emph {et~al.}(2018)\citenamefont {Cotner},
  \citenamefont {Kusenko},\ and\ \citenamefont {Takhistov}}]{Cotner:2018vug}%
  \BibitemOpen
  \bibfield  {author} {\bibinfo {author} {\bibfnamefont {E.}~\bibnamefont
  {Cotner}}, \bibinfo {author} {\bibfnamefont {A.}~\bibnamefont {Kusenko}}, \
  and\ \bibinfo {author} {\bibfnamefont {V.}~\bibnamefont {Takhistov}},\ }\href
  {\doibase 10.1103/PhysRevD.98.083513} {\bibfield  {journal} {\bibinfo
  {journal} {Phys. Rev.}\ }\textbf {\bibinfo {volume} {D98}},\ \bibinfo {pages}
  {083513} (\bibinfo {year} {2018})},\ \Eprint
  {http://arxiv.org/abs/1801.03321} {arXiv:1801.03321 [astro-ph.CO]}
  \BibitemShut {NoStop}%
\bibitem [{\citenamefont {Sasaki}\ \emph {et~al.}(2018)\citenamefont {Sasaki},
  \citenamefont {Suyama}, \citenamefont {Tanaka},\ and\ \citenamefont
  {Yokoyama}}]{Sasaki:2018dmp}%
  \BibitemOpen
  \bibfield  {author} {\bibinfo {author} {\bibfnamefont {M.}~\bibnamefont
  {Sasaki}}, \bibinfo {author} {\bibfnamefont {T.}~\bibnamefont {Suyama}},
  \bibinfo {author} {\bibfnamefont {T.}~\bibnamefont {Tanaka}}, \ and\ \bibinfo
  {author} {\bibfnamefont {S.}~\bibnamefont {Yokoyama}},\ }\href {\doibase
  10.1088/1361-6382/aaa7b4} {\bibfield  {journal} {\bibinfo  {journal} {Class.
  Quant. Grav.}\ }\textbf {\bibinfo {volume} {35}},\ \bibinfo {pages} {063001}
  (\bibinfo {year} {2018})},\ \Eprint {http://arxiv.org/abs/1801.05235}
  {arXiv:1801.05235 [astro-ph.CO]} \BibitemShut {NoStop}%
\bibitem [{\citenamefont {Carr}\ and\ \citenamefont
  {Silk}(2018)}]{Carr:2018rid}%
  \BibitemOpen
  \bibfield  {author} {\bibinfo {author} {\bibfnamefont {B.}~\bibnamefont
  {Carr}}\ and\ \bibinfo {author} {\bibfnamefont {J.}~\bibnamefont {Silk}},\
  }\href {\doibase 10.1093/mnras/sty1204} {\bibfield  {journal} {\bibinfo
  {journal} {Mon. Not. Roy. Astron. Soc.}\ }\textbf {\bibinfo {volume} {478}},\
  \bibinfo {pages} {3756} (\bibinfo {year} {2018})},\ \Eprint
  {http://arxiv.org/abs/1801.00672} {arXiv:1801.00672 [astro-ph.CO]}
  \BibitemShut {NoStop}%
\bibitem [{\citenamefont {Cotner}\ \emph {et~al.}(2019)\citenamefont {Cotner},
  \citenamefont {Kusenko}, \citenamefont {Sasaki},\ and\ \citenamefont
  {Takhistov}}]{Cotner:2019ykd}%
  \BibitemOpen
  \bibfield  {author} {\bibinfo {author} {\bibfnamefont {E.}~\bibnamefont
  {Cotner}}, \bibinfo {author} {\bibfnamefont {A.}~\bibnamefont {Kusenko}},
  \bibinfo {author} {\bibfnamefont {M.}~\bibnamefont {Sasaki}}, \ and\ \bibinfo
  {author} {\bibfnamefont {V.}~\bibnamefont {Takhistov}},\ }\href {\doibase
  10.1088/1475-7516/2019/10/077} {\bibfield  {journal} {\bibinfo  {journal}
  {JCAP}\ }\textbf {\bibinfo {volume} {1910}},\ \bibinfo {pages} {077}
  (\bibinfo {year} {2019})},\ \Eprint {http://arxiv.org/abs/1907.10613}
  {arXiv:1907.10613 [astro-ph.CO]} \BibitemShut {NoStop}%
\bibitem [{\citenamefont {Nakamura}\ \emph {et~al.}(1997)\citenamefont
  {Nakamura}, \citenamefont {Sasaki}, \citenamefont {Tanaka},\ and\
  \citenamefont {Thorne}}]{Nakamura:1997sm}%
  \BibitemOpen
  \bibfield  {author} {\bibinfo {author} {\bibfnamefont {T.}~\bibnamefont
  {Nakamura}}, \bibinfo {author} {\bibfnamefont {M.}~\bibnamefont {Sasaki}},
  \bibinfo {author} {\bibfnamefont {T.}~\bibnamefont {Tanaka}}, \ and\ \bibinfo
  {author} {\bibfnamefont {K.~S.}\ \bibnamefont {Thorne}},\ }\href {\doibase
  10.1086/310886} {\bibfield  {journal} {\bibinfo  {journal} {Astrophys. J.}\
  }\textbf {\bibinfo {volume} {487}},\ \bibinfo {pages} {L139} (\bibinfo {year}
  {1997})},\ \Eprint {http://arxiv.org/abs/astro-ph/9708060}
  {arXiv:astro-ph/9708060 [astro-ph]} \BibitemShut {NoStop}%
\bibitem [{\citenamefont {Clesse}\ and\ \citenamefont
  {Garcia-Bellido}(2015)}]{Clesse:2015wea}%
  \BibitemOpen
  \bibfield  {author} {\bibinfo {author} {\bibfnamefont {S.}~\bibnamefont
  {Clesse}}\ and\ \bibinfo {author} {\bibfnamefont {J.}~\bibnamefont
  {Garcia-Bellido}},\ }\href {\doibase 10.1103/PhysRevD.92.023524} {\bibfield
  {journal} {\bibinfo  {journal} {Phys. Rev.}\ }\textbf {\bibinfo {volume}
  {D92}},\ \bibinfo {pages} {023524} (\bibinfo {year} {2015})},\ \Eprint
  {http://arxiv.org/abs/1501.07565} {arXiv:1501.07565 [astro-ph.CO]}
  \BibitemShut {NoStop}%
\bibitem [{\citenamefont {Bird}\ \emph {et~al.}(2016)\citenamefont {Bird} \emph
  {et~al.}}]{Bird:2016dcv}%
  \BibitemOpen
  \bibfield  {author} {\bibinfo {author} {\bibfnamefont {S.}~\bibnamefont
  {Bird}} \emph {et~al.},\ }\href {\doibase 10.1103/PhysRevLett.116.201301}
  {\bibfield  {journal} {\bibinfo  {journal} {Phys. Rev. Lett.}\ }\textbf
  {\bibinfo {volume} {116}},\ \bibinfo {pages} {201301} (\bibinfo {year}
  {2016})},\ \Eprint {http://arxiv.org/abs/1603.00464} {arXiv:1603.00464
  [astro-ph.CO]} \BibitemShut {NoStop}%
\bibitem [{\citenamefont {Raidal}\ \emph {et~al.}(2017)\citenamefont {Raidal},
  \citenamefont {Vaskonen},\ and\ \citenamefont {Veerm\"ae}}]{Raidal:2017mfl}%
  \BibitemOpen
  \bibfield  {author} {\bibinfo {author} {\bibfnamefont {M.}~\bibnamefont
  {Raidal}}, \bibinfo {author} {\bibfnamefont {V.}~\bibnamefont {Vaskonen}}, \
  and\ \bibinfo {author} {\bibfnamefont {H.}~\bibnamefont {Veerm\"ae}},\ }\href
  {\doibase 10.1088/1475-7516/2017/09/037} {\bibfield  {journal} {\bibinfo
  {journal} {JCAP}\ }\textbf {\bibinfo {volume} {09}},\ \bibinfo {pages} {037}
  (\bibinfo {year} {2017})},\ \Eprint {http://arxiv.org/abs/1707.01480}
  {arXiv:1707.01480 [astro-ph.CO]} \BibitemShut {NoStop}%
\bibitem [{\citenamefont {Eroshenko}(2018)}]{Eroshenko:2016hmn}%
  \BibitemOpen
  \bibfield  {author} {\bibinfo {author} {\bibfnamefont {Y.~N.}\ \bibnamefont
  {Eroshenko}},\ }\href {\doibase 10.1088/1742-6596/1051/1/012010} {\bibfield
  {journal} {\bibinfo  {journal} {J. Phys. Conf. Ser.}\ }\textbf {\bibinfo
  {volume} {1051}},\ \bibinfo {pages} {012010} (\bibinfo {year} {2018})},\
  \Eprint {http://arxiv.org/abs/1604.04932} {arXiv:1604.04932 [astro-ph.CO]}
  \BibitemShut {NoStop}%
\bibitem [{\citenamefont {Sasaki}\ \emph {et~al.}(2016)\citenamefont {Sasaki},
  \citenamefont {Suyama}, \citenamefont {Tanaka},\ and\ \citenamefont
  {Yokoyama}}]{Sasaki:2016jop}%
  \BibitemOpen
  \bibfield  {author} {\bibinfo {author} {\bibfnamefont {M.}~\bibnamefont
  {Sasaki}}, \bibinfo {author} {\bibfnamefont {T.}~\bibnamefont {Suyama}},
  \bibinfo {author} {\bibfnamefont {T.}~\bibnamefont {Tanaka}}, \ and\ \bibinfo
  {author} {\bibfnamefont {S.}~\bibnamefont {Yokoyama}},\ }\href {\doibase
  10.1103/PhysRevLett.117.061101} {\bibfield  {journal} {\bibinfo  {journal}
  {Phys. Rev. Lett.}\ }\textbf {\bibinfo {volume} {117}},\ \bibinfo {pages}
  {061101} (\bibinfo {year} {2016})},\ \Eprint
  {http://arxiv.org/abs/1603.08338} {arXiv:1603.08338 [astro-ph.CO]}
  \BibitemShut {NoStop}%
\bibitem [{\citenamefont {Clesse}\ and\ \citenamefont
  {Garc\'\i{}a-Bellido}(2017)}]{Clesse:2016ajp}%
  \BibitemOpen
  \bibfield  {author} {\bibinfo {author} {\bibfnamefont {S.}~\bibnamefont
  {Clesse}}\ and\ \bibinfo {author} {\bibfnamefont {J.}~\bibnamefont
  {Garc\'\i{}a-Bellido}},\ }\href {\doibase 10.1016/j.dark.2017.10.001}
  {\bibfield  {journal} {\bibinfo  {journal} {Phys. Dark Univ.}\ }\textbf
  {\bibinfo {volume} {18}},\ \bibinfo {pages} {105} (\bibinfo {year} {2017})},\
  \Eprint {http://arxiv.org/abs/1610.08479} {arXiv:1610.08479 [astro-ph.CO]}
  \BibitemShut {NoStop}%
\bibitem [{\citenamefont {Fuller}\ \emph {et~al.}(2017)\citenamefont {Fuller},
  \citenamefont {Kusenko},\ and\ \citenamefont {Takhistov}}]{Fuller:2017uyd}%
  \BibitemOpen
  \bibfield  {author} {\bibinfo {author} {\bibfnamefont {G.~M.}\ \bibnamefont
  {Fuller}}, \bibinfo {author} {\bibfnamefont {A.}~\bibnamefont {Kusenko}}, \
  and\ \bibinfo {author} {\bibfnamefont {V.}~\bibnamefont {Takhistov}},\ }\href
  {\doibase 10.1103/PhysRevLett.119.061101} {\bibfield  {journal} {\bibinfo
  {journal} {Phys. Rev. Lett.}\ }\textbf {\bibinfo {volume} {119}},\ \bibinfo
  {pages} {061101} (\bibinfo {year} {2017})},\ \Eprint
  {http://arxiv.org/abs/1704.01129} {arXiv:1704.01129 [astro-ph.HE]}
  \BibitemShut {NoStop}%
\bibitem [{\citenamefont {Abbott}\ \emph
  {et~al.}(2016{\natexlab{a}})\citenamefont {Abbott} \emph
  {et~al.}}]{Abbott:2016blz}%
  \BibitemOpen
  \bibfield  {author} {\bibinfo {author} {\bibfnamefont {B.~P.}\ \bibnamefont
  {Abbott}} \emph {et~al.} (\bibinfo {collaboration} {Virgo, LIGO
  Scientific}),\ }\href {\doibase 10.1103/PhysRevLett.116.061102} {\bibfield
  {journal} {\bibinfo  {journal} {Phys. Rev. Lett.}\ }\textbf {\bibinfo
  {volume} {116}},\ \bibinfo {pages} {061102} (\bibinfo {year}
  {2016}{\natexlab{a}})},\ \Eprint {http://arxiv.org/abs/1602.03837}
  {arXiv:1602.03837 [gr-qc]} \BibitemShut {NoStop}%
\bibitem [{\citenamefont {Abbott}\ \emph
  {et~al.}(2016{\natexlab{b}})\citenamefont {Abbott} \emph
  {et~al.}}]{Abbott:2016nmj}%
  \BibitemOpen
  \bibfield  {author} {\bibinfo {author} {\bibfnamefont {B.~P.}\ \bibnamefont
  {Abbott}} \emph {et~al.} (\bibinfo {collaboration} {Virgo, LIGO
  Scientific}),\ }\href {\doibase 10.1103/PhysRevLett.116.241103} {\bibfield
  {journal} {\bibinfo  {journal} {Phys. Rev. Lett.}\ }\textbf {\bibinfo
  {volume} {116}},\ \bibinfo {pages} {241103} (\bibinfo {year}
  {2016}{\natexlab{b}})},\ \Eprint {http://arxiv.org/abs/1606.04855}
  {arXiv:1606.04855 [gr-qc]} \BibitemShut {NoStop}%
\bibitem [{\citenamefont {Abbott}\ \emph {et~al.}(2017)\citenamefont {Abbott}
  \emph {et~al.}}]{Abbott:2017vtc}%
  \BibitemOpen
  \bibfield  {author} {\bibinfo {author} {\bibfnamefont {B.~P.}\ \bibnamefont
  {Abbott}} \emph {et~al.} (\bibinfo {collaboration} {VIRGO, LIGO
  Scientific}),\ }\href {\doibase 10.1103/PhysRevLett.118.221101} {\bibfield
  {journal} {\bibinfo  {journal} {Phys. Rev. Lett.}\ }\textbf {\bibinfo
  {volume} {118}},\ \bibinfo {pages} {221101} (\bibinfo {year} {2017})},\
  \Eprint {http://arxiv.org/abs/1706.01812} {arXiv:1706.01812 [gr-qc]}
  \BibitemShut {NoStop}%
\bibitem [{\citenamefont {Bean}\ and\ \citenamefont
  {Magueijo}(2002)}]{Bean:2002kx}%
  \BibitemOpen
  \bibfield  {author} {\bibinfo {author} {\bibfnamefont {R.}~\bibnamefont
  {Bean}}\ and\ \bibinfo {author} {\bibfnamefont {J.}~\bibnamefont
  {Magueijo}},\ }\href {\doibase 10.1103/PhysRevD.66.063505} {\bibfield
  {journal} {\bibinfo  {journal} {Phys. Rev.}\ }\textbf {\bibinfo {volume}
  {D66}},\ \bibinfo {pages} {063505} (\bibinfo {year} {2002})},\ \Eprint
  {http://arxiv.org/abs/astro-ph/0204486} {arXiv:astro-ph/0204486 [astro-ph]}
  \BibitemShut {NoStop}%
\bibitem [{\citenamefont {Kawasaki}\ \emph {et~al.}(2012)\citenamefont
  {Kawasaki}, \citenamefont {Kusenko},\ and\ \citenamefont
  {Yanagida}}]{Kawasaki:2012kn}%
  \BibitemOpen
  \bibfield  {author} {\bibinfo {author} {\bibfnamefont {M.}~\bibnamefont
  {Kawasaki}}, \bibinfo {author} {\bibfnamefont {A.}~\bibnamefont {Kusenko}}, \
  and\ \bibinfo {author} {\bibfnamefont {T.~T.}\ \bibnamefont {Yanagida}},\
  }\href {\doibase 10.1016/j.physletb.2012.03.056} {\bibfield  {journal}
  {\bibinfo  {journal} {Phys. Lett.}\ }\textbf {\bibinfo {volume} {B711}},\
  \bibinfo {pages} {1} (\bibinfo {year} {2012})},\ \Eprint
  {http://arxiv.org/abs/1202.3848} {arXiv:1202.3848 [astro-ph.CO]} \BibitemShut
  {NoStop}%
\bibitem [{\citenamefont {Takhistov}(2019)}]{Takhistov:2017nmt}%
  \BibitemOpen
  \bibfield  {author} {\bibinfo {author} {\bibfnamefont {V.}~\bibnamefont
  {Takhistov}},\ }\href {\doibase 10.1016/j.physletb.2018.12.043} {\bibfield
  {journal} {\bibinfo  {journal} {Phys. Lett.}\ }\textbf {\bibinfo {volume}
  {B789}},\ \bibinfo {pages} {538} (\bibinfo {year} {2019})},\ \Eprint
  {http://arxiv.org/abs/1710.09458} {arXiv:1710.09458 [astro-ph.HE]}
  \BibitemShut {NoStop}%
\bibitem [{\citenamefont {Takhistov}(2018)}]{Takhistov:2017bpt}%
  \BibitemOpen
  \bibfield  {author} {\bibinfo {author} {\bibfnamefont {V.}~\bibnamefont
  {Takhistov}},\ }\href {\doibase 10.1016/j.physletb.2018.05.026} {\bibfield
  {journal} {\bibinfo  {journal} {Phys. Lett.}\ }\textbf {\bibinfo {volume}
  {B782}},\ \bibinfo {pages} {77} (\bibinfo {year} {2018})},\ \Eprint
  {http://arxiv.org/abs/1707.05849} {arXiv:1707.05849 [astro-ph.CO]}
  \BibitemShut {NoStop}%
\bibitem [{\citenamefont {Caldwell}\ and\ \citenamefont
  {Casper}(1996)}]{Caldwell:1995fu}%
  \BibitemOpen
  \bibfield  {author} {\bibinfo {author} {\bibfnamefont {R.~R.}\ \bibnamefont
  {Caldwell}}\ and\ \bibinfo {author} {\bibfnamefont {P.}~\bibnamefont
  {Casper}},\ }\href {\doibase 10.1103/PhysRevD.53.3002} {\bibfield  {journal}
  {\bibinfo  {journal} {Phys. Rev.}\ }\textbf {\bibinfo {volume} {D53}},\
  \bibinfo {pages} {3002} (\bibinfo {year} {1996})},\ \Eprint
  {http://arxiv.org/abs/gr-qc/9509012} {arXiv:gr-qc/9509012 [gr-qc]}
  \BibitemShut {NoStop}%
\bibitem [{\citenamefont {Garriga}\ and\ \citenamefont
  {Vilenkin}(1993)}]{Garriga:1992nm}%
  \BibitemOpen
  \bibfield  {author} {\bibinfo {author} {\bibfnamefont {J.}~\bibnamefont
  {Garriga}}\ and\ \bibinfo {author} {\bibfnamefont {A.}~\bibnamefont
  {Vilenkin}},\ }\href {\doibase 10.1103/PhysRevD.47.3265} {\bibfield
  {journal} {\bibinfo  {journal} {Phys. Rev.}\ }\textbf {\bibinfo {volume}
  {D47}},\ \bibinfo {pages} {3265} (\bibinfo {year} {1993})},\ \Eprint
  {http://arxiv.org/abs/hep-ph/9208212} {arXiv:hep-ph/9208212 [hep-ph]}
  \BibitemShut {NoStop}%
\bibitem [{\citenamefont {Hawking}\ \emph {et~al.}(1982)\citenamefont
  {Hawking}, \citenamefont {Moss},\ and\ \citenamefont
  {Stewart}}]{Hawking:1982ga}%
  \BibitemOpen
  \bibfield  {author} {\bibinfo {author} {\bibfnamefont {S.~W.}\ \bibnamefont
  {Hawking}}, \bibinfo {author} {\bibfnamefont {I.~G.}\ \bibnamefont {Moss}}, \
  and\ \bibinfo {author} {\bibfnamefont {J.~M.}\ \bibnamefont {Stewart}},\
  }\href {\doibase 10.1103/PhysRevD.26.2681} {\bibfield  {journal} {\bibinfo
  {journal} {Phys. Rev.}\ }\textbf {\bibinfo {volume} {D26}},\ \bibinfo {pages}
  {2681} (\bibinfo {year} {1982})}\BibitemShut {NoStop}%
\bibitem [{\citenamefont {Lewicki}\ and\ \citenamefont
  {Vaskonen}(2020)}]{Lewicki:2019gmv}%
  \BibitemOpen
  \bibfield  {author} {\bibinfo {author} {\bibfnamefont {M.}~\bibnamefont
  {Lewicki}}\ and\ \bibinfo {author} {\bibfnamefont {V.}~\bibnamefont
  {Vaskonen}},\ }\href {\doibase 10.1016/j.dark.2020.100672} {\bibfield
  {journal} {\bibinfo  {journal} {Phys. Dark Univ.}\ }\textbf {\bibinfo
  {volume} {30}},\ \bibinfo {pages} {100672} (\bibinfo {year} {2020})},\
  \Eprint {http://arxiv.org/abs/1912.00997} {arXiv:1912.00997 [astro-ph.CO]}
  \BibitemShut {NoStop}%
\bibitem [{\citenamefont {Deng}\ \emph {et~al.}(2017)\citenamefont {Deng},
  \citenamefont {Garriga},\ and\ \citenamefont {Vilenkin}}]{Deng:2016vzb}%
  \BibitemOpen
  \bibfield  {author} {\bibinfo {author} {\bibfnamefont {H.}~\bibnamefont
  {Deng}}, \bibinfo {author} {\bibfnamefont {J.}~\bibnamefont {Garriga}}, \
  and\ \bibinfo {author} {\bibfnamefont {A.}~\bibnamefont {Vilenkin}},\ }\href
  {\doibase 10.1088/1475-7516/2017/04/050} {\bibfield  {journal} {\bibinfo
  {journal} {JCAP}\ }\textbf {\bibinfo {volume} {1704}},\ \bibinfo {pages}
  {050} (\bibinfo {year} {2017})},\ \Eprint {http://arxiv.org/abs/1612.03753}
  {arXiv:1612.03753 [gr-qc]} \BibitemShut {NoStop}%
\bibitem [{\citenamefont {Katz}\ \emph {et~al.}(2018)\citenamefont {Katz},
  \citenamefont {Kopp}, \citenamefont {Sibiryakov},\ and\ \citenamefont
  {Xue}}]{Katz:2018zrn}%
  \BibitemOpen
  \bibfield  {author} {\bibinfo {author} {\bibfnamefont {A.}~\bibnamefont
  {Katz}}, \bibinfo {author} {\bibfnamefont {J.}~\bibnamefont {Kopp}}, \bibinfo
  {author} {\bibfnamefont {S.}~\bibnamefont {Sibiryakov}}, \ and\ \bibinfo
  {author} {\bibfnamefont {W.}~\bibnamefont {Xue}},\ }\href {\doibase
  10.1088/1475-7516/2018/12/005} {\bibfield  {journal} {\bibinfo  {journal}
  {JCAP}\ }\textbf {\bibinfo {volume} {12}},\ \bibinfo {pages} {005} (\bibinfo
  {year} {2018})},\ \Eprint {http://arxiv.org/abs/1807.11495} {arXiv:1807.11495
  [astro-ph.CO]} \BibitemShut {NoStop}%
\bibitem [{\citenamefont {Smyth}\ \emph {et~al.}(2020)\citenamefont {Smyth},
  \citenamefont {Profumo}, \citenamefont {English}, \citenamefont {Jeltema},
  \citenamefont {McKinnon},\ and\ \citenamefont
  {Guhathakurta}}]{Smyth:2019whb}%
  \BibitemOpen
  \bibfield  {author} {\bibinfo {author} {\bibfnamefont {N.}~\bibnamefont
  {Smyth}}, \bibinfo {author} {\bibfnamefont {S.}~\bibnamefont {Profumo}},
  \bibinfo {author} {\bibfnamefont {S.}~\bibnamefont {English}}, \bibinfo
  {author} {\bibfnamefont {T.}~\bibnamefont {Jeltema}}, \bibinfo {author}
  {\bibfnamefont {K.}~\bibnamefont {McKinnon}}, \ and\ \bibinfo {author}
  {\bibfnamefont {P.}~\bibnamefont {Guhathakurta}},\ }\href {\doibase
  10.1103/PhysRevD.101.063005} {\bibfield  {journal} {\bibinfo  {journal}
  {Phys. Rev. D}\ }\textbf {\bibinfo {volume} {101}},\ \bibinfo {pages}
  {063005} (\bibinfo {year} {2020})},\ \Eprint
  {http://arxiv.org/abs/1910.01285} {arXiv:1910.01285 [astro-ph.CO]}
  \BibitemShut {NoStop}%
\bibitem [{\citenamefont {Montero-Camacho}\ \emph {et~al.}(2019)\citenamefont
  {Montero-Camacho}, \citenamefont {Fang}, \citenamefont {Vasquez},
  \citenamefont {Silva},\ and\ \citenamefont
  {Hirata}}]{Montero-Camacho:2019jte}%
  \BibitemOpen
  \bibfield  {author} {\bibinfo {author} {\bibfnamefont {P.}~\bibnamefont
  {Montero-Camacho}}, \bibinfo {author} {\bibfnamefont {X.}~\bibnamefont
  {Fang}}, \bibinfo {author} {\bibfnamefont {G.}~\bibnamefont {Vasquez}},
  \bibinfo {author} {\bibfnamefont {M.}~\bibnamefont {Silva}}, \ and\ \bibinfo
  {author} {\bibfnamefont {C.~M.}\ \bibnamefont {Hirata}},\ }\href {\doibase
  10.1088/1475-7516/2019/08/031} {\bibfield  {journal} {\bibinfo  {journal}
  {JCAP}\ }\textbf {\bibinfo {volume} {1908}},\ \bibinfo {pages} {031}
  (\bibinfo {year} {2019})},\ \Eprint {http://arxiv.org/abs/1906.05950}
  {arXiv:1906.05950 [astro-ph.CO]} \BibitemShut {NoStop}%
\bibitem [{\citenamefont {Garriga}\ \emph {et~al.}(2016)\citenamefont
  {Garriga}, \citenamefont {Vilenkin},\ and\ \citenamefont
  {Zhang}}]{Garriga:2015fdk}%
  \BibitemOpen
  \bibfield  {author} {\bibinfo {author} {\bibfnamefont {J.}~\bibnamefont
  {Garriga}}, \bibinfo {author} {\bibfnamefont {A.}~\bibnamefont {Vilenkin}}, \
  and\ \bibinfo {author} {\bibfnamefont {J.}~\bibnamefont {Zhang}},\ }\href
  {\doibase 10.1088/1475-7516/2016/02/064} {\bibfield  {journal} {\bibinfo
  {journal} {JCAP}\ }\textbf {\bibinfo {volume} {1602}},\ \bibinfo {pages}
  {064} (\bibinfo {year} {2016})},\ \Eprint {http://arxiv.org/abs/1512.01819}
  {arXiv:1512.01819 [hep-th]} \BibitemShut {NoStop}%
\bibitem [{\citenamefont {Deng}\ and\ \citenamefont
  {Vilenkin}(2017)}]{Deng:2017uwc}%
  \BibitemOpen
  \bibfield  {author} {\bibinfo {author} {\bibfnamefont {H.}~\bibnamefont
  {Deng}}\ and\ \bibinfo {author} {\bibfnamefont {A.}~\bibnamefont
  {Vilenkin}},\ }\href {\doibase 10.1088/1475-7516/2017/12/044} {\bibfield
  {journal} {\bibinfo  {journal} {JCAP}\ }\textbf {\bibinfo {volume} {1712}},\
  \bibinfo {pages} {044} (\bibinfo {year} {2017})},\ \Eprint
  {http://arxiv.org/abs/1710.02865} {arXiv:1710.02865 [gr-qc]} \BibitemShut
  {NoStop}%
\bibitem [{\citenamefont {Deng}\ \emph {et~al.}(2018)\citenamefont {Deng},
  \citenamefont {Vilenkin},\ and\ \citenamefont {Yamada}}]{Deng:2018cxb}%
  \BibitemOpen
  \bibfield  {author} {\bibinfo {author} {\bibfnamefont {H.}~\bibnamefont
  {Deng}}, \bibinfo {author} {\bibfnamefont {A.}~\bibnamefont {Vilenkin}}, \
  and\ \bibinfo {author} {\bibfnamefont {M.}~\bibnamefont {Yamada}},\ }\href
  {\doibase 10.1088/1475-7516/2018/07/059} {\bibfield  {journal} {\bibinfo
  {journal} {JCAP}\ }\textbf {\bibinfo {volume} {1807}},\ \bibinfo {pages}
  {059} (\bibinfo {year} {2018})},\ \Eprint {http://arxiv.org/abs/1804.10059}
  {arXiv:1804.10059 [gr-qc]} \BibitemShut {NoStop}%
\bibitem [{\citenamefont {Niikura}\ \emph
  {et~al.}(2019{\natexlab{a}})\citenamefont {Niikura} \emph
  {et~al.}}]{Niikura:2017zjd}%
  \BibitemOpen
  \bibfield  {author} {\bibinfo {author} {\bibfnamefont {H.}~\bibnamefont
  {Niikura}} \emph {et~al.},\ }\href {\doibase 10.1038/s41550-019-0723-1}
  {\bibfield  {journal} {\bibinfo  {journal} {Nat. Astron.}\ }\textbf {\bibinfo
  {volume} {3}},\ \bibinfo {pages} {524} (\bibinfo {year}
  {2019}{\natexlab{a}})},\ \Eprint {http://arxiv.org/abs/1701.02151}
  {arXiv:1701.02151 [astro-ph.CO]} \BibitemShut {NoStop}%
\bibitem [{\citenamefont {Susskind}(2003)}]{Susskind:2003kw}%
  \BibitemOpen
  \bibfield  {author} {\bibinfo {author} {\bibfnamefont {L.}~\bibnamefont
  {Susskind}},\ }\href@noop {} {\ ,\ \bibinfo {pages} {247} (\bibinfo {year}
  {2003})},\ \Eprint {http://arxiv.org/abs/hep-th/0302219}
  {arXiv:hep-th/0302219 [hep-th]} \BibitemShut {NoStop}%
\bibitem [{\citenamefont {Coleman}\ and\ \citenamefont
  {De~Luccia}(1980)}]{Coleman:1980aw}%
  \BibitemOpen
  \bibfield  {author} {\bibinfo {author} {\bibfnamefont {S.~R.}\ \bibnamefont
  {Coleman}}\ and\ \bibinfo {author} {\bibfnamefont {F.}~\bibnamefont
  {De~Luccia}},\ }\href {\doibase 10.1103/PhysRevD.21.3305} {\bibfield
  {journal} {\bibinfo  {journal} {Phys. Rev.}\ }\textbf {\bibinfo {volume}
  {D21}},\ \bibinfo {pages} {3305} (\bibinfo {year} {1980})}\BibitemShut
  {NoStop}%
\bibitem [{\citenamefont {Baumann}\ and\ \citenamefont
  {McAllister}(2015)}]{Baumann:2014nda}%
  \BibitemOpen
  \bibfield  {author} {\bibinfo {author} {\bibfnamefont {D.}~\bibnamefont
  {Baumann}}\ and\ \bibinfo {author} {\bibfnamefont {L.}~\bibnamefont
  {McAllister}},\ }\href {\doibase 10.1017/CBO9781316105733} {\emph {\bibinfo
  {title} {{Inflation and String Theory}}}},\ Cambridge Monographs on
  Mathematical Physics\ (\bibinfo  {publisher} {Cambridge University Press},\
  \bibinfo {year} {2015})\ \Eprint {http://arxiv.org/abs/1404.2601}
  {arXiv:1404.2601 [hep-th]} \BibitemShut {NoStop}%
\bibitem [{\citenamefont {Lee}\ and\ \citenamefont
  {Weinberg}(1987)}]{Lee:1987qc}%
  \BibitemOpen
  \bibfield  {author} {\bibinfo {author} {\bibfnamefont {K.-M.}\ \bibnamefont
  {Lee}}\ and\ \bibinfo {author} {\bibfnamefont {E.~J.}\ \bibnamefont
  {Weinberg}},\ }\href {\doibase 10.1103/PhysRevD.36.1088} {\bibfield
  {journal} {\bibinfo  {journal} {Phys. Rev.}\ }\textbf {\bibinfo {volume}
  {D36}},\ \bibinfo {pages} {1088} (\bibinfo {year} {1987})}\BibitemShut
  {NoStop}%
\bibitem [{\citenamefont {Coleman}(1985)}]{Coleman:1985ki}%
  \BibitemOpen
  \bibfield  {author} {\bibinfo {author} {\bibfnamefont {S.~R.}\ \bibnamefont
  {Coleman}},\ }\href {\doibase 10.1016/0550-3213(85)90286-X,
  10.1016/0550-3213(86)90520-1} {\bibfield  {journal} {\bibinfo  {journal}
  {Nucl. Phys.}\ }\textbf {\bibinfo {volume} {B262}},\ \bibinfo {pages} {263}
  (\bibinfo {year} {1985})},\ \bibinfo {note} {[Erratum: Nucl.
  Phys.B269,744(1986)]}\BibitemShut {NoStop}%
\bibitem [{\citenamefont {Kusenko}(1995)}]{Kusenko:1995jv}%
  \BibitemOpen
  \bibfield  {author} {\bibinfo {author} {\bibfnamefont {A.}~\bibnamefont
  {Kusenko}},\ }\href {\doibase 10.1016/0370-2693(95)00994-V} {\bibfield
  {journal} {\bibinfo  {journal} {Phys. Lett. B}\ }\textbf {\bibinfo {volume}
  {358}},\ \bibinfo {pages} {51} (\bibinfo {year} {1995})},\ \Eprint
  {http://arxiv.org/abs/hep-ph/9504418} {arXiv:hep-ph/9504418} \BibitemShut
  {NoStop}%
\bibitem [{\citenamefont {Blau}\ \emph {et~al.}(1987)\citenamefont {Blau},
  \citenamefont {Guendelman},\ and\ \citenamefont {Guth}}]{Blau:1986cw}%
  \BibitemOpen
  \bibfield  {author} {\bibinfo {author} {\bibfnamefont {S.~K.}\ \bibnamefont
  {Blau}}, \bibinfo {author} {\bibfnamefont {E.~I.}\ \bibnamefont
  {Guendelman}}, \ and\ \bibinfo {author} {\bibfnamefont {A.~H.}\ \bibnamefont
  {Guth}},\ }\href {\doibase 10.1103/PhysRevD.35.1747} {\bibfield  {journal}
  {\bibinfo  {journal} {Phys. Rev.}\ }\textbf {\bibinfo {volume} {D35}},\
  \bibinfo {pages} {1747} (\bibinfo {year} {1987})}\BibitemShut {NoStop}%
\bibitem [{\citenamefont {Maeda}\ \emph {et~al.}(1982)\citenamefont {Maeda},
  \citenamefont {Sato}, \citenamefont {Sasaki},\ and\ \citenamefont
  {Kodama}}]{Maeda:1981gw}%
  \BibitemOpen
  \bibfield  {author} {\bibinfo {author} {\bibfnamefont {K.-i.}\ \bibnamefont
  {Maeda}}, \bibinfo {author} {\bibfnamefont {K.}~\bibnamefont {Sato}},
  \bibinfo {author} {\bibfnamefont {M.}~\bibnamefont {Sasaki}}, \ and\ \bibinfo
  {author} {\bibfnamefont {H.}~\bibnamefont {Kodama}},\ }\href {\doibase
  10.1016/0370-2693(82)91151-0} {\bibfield  {journal} {\bibinfo  {journal}
  {Phys. Lett.}\ }\textbf {\bibinfo {volume} {108B}},\ \bibinfo {pages} {98}
  (\bibinfo {year} {1982})}\BibitemShut {NoStop}%
\bibitem [{\citenamefont {Kodama}\ \emph {et~al.}(1981)\citenamefont {Kodama},
  \citenamefont {Sasaki}, \citenamefont {Sato},\ and\ \citenamefont
  {Maeda}}]{Kodama:1981gu}%
  \BibitemOpen
  \bibfield  {author} {\bibinfo {author} {\bibfnamefont {H.}~\bibnamefont
  {Kodama}}, \bibinfo {author} {\bibfnamefont {M.}~\bibnamefont {Sasaki}},
  \bibinfo {author} {\bibfnamefont {K.}~\bibnamefont {Sato}}, \ and\ \bibinfo
  {author} {\bibfnamefont {K.-i.}\ \bibnamefont {Maeda}},\ }\href {\doibase
  10.1143/PTP.66.2052} {\bibfield  {journal} {\bibinfo  {journal} {Prog. Theor.
  Phys.}\ }\textbf {\bibinfo {volume} {66}},\ \bibinfo {pages} {2052} (\bibinfo
  {year} {1981})}\BibitemShut {NoStop}%
\bibitem [{\citenamefont {Sato}\ \emph {et~al.}(1981)\citenamefont {Sato},
  \citenamefont {Sasaki}, \citenamefont {Kodama},\ and\ \citenamefont
  {Maeda}}]{Sato:1981bf}%
  \BibitemOpen
  \bibfield  {author} {\bibinfo {author} {\bibfnamefont {K.}~\bibnamefont
  {Sato}}, \bibinfo {author} {\bibfnamefont {M.}~\bibnamefont {Sasaki}},
  \bibinfo {author} {\bibfnamefont {H.}~\bibnamefont {Kodama}}, \ and\ \bibinfo
  {author} {\bibfnamefont {K.-i.}\ \bibnamefont {Maeda}},\ }\href {\doibase
  10.1143/PTP.65.1443} {\bibfield  {journal} {\bibinfo  {journal} {Prog. Theor.
  Phys.}\ }\textbf {\bibinfo {volume} {65}},\ \bibinfo {pages} {1443} (\bibinfo
  {year} {1981})}\BibitemShut {NoStop}%
\bibitem [{\citenamefont {Sato}\ \emph {et~al.}(1982)\citenamefont {Sato},
  \citenamefont {Kodama}, \citenamefont {Sasaki},\ and\ \citenamefont
  {Maeda}}]{Sato:1981gv}%
  \BibitemOpen
  \bibfield  {author} {\bibinfo {author} {\bibfnamefont {K.}~\bibnamefont
  {Sato}}, \bibinfo {author} {\bibfnamefont {H.}~\bibnamefont {Kodama}},
  \bibinfo {author} {\bibfnamefont {M.}~\bibnamefont {Sasaki}}, \ and\ \bibinfo
  {author} {\bibfnamefont {K.-i.}\ \bibnamefont {Maeda}},\ }\href {\doibase
  10.1016/0370-2693(82)91152-2} {\bibfield  {journal} {\bibinfo  {journal}
  {Phys. Lett.}\ }\textbf {\bibinfo {volume} {108B}},\ \bibinfo {pages} {103}
  (\bibinfo {year} {1982})}\BibitemShut {NoStop}%
\bibitem [{\citenamefont {Linde}(2017)}]{Linde:2015edk}%
  \BibitemOpen
  \bibfield  {author} {\bibinfo {author} {\bibfnamefont {A.}~\bibnamefont
  {Linde}},\ }\href {\doibase 10.1088/1361-6633/aa50e4} {\bibfield  {journal}
  {\bibinfo  {journal} {Rept. Prog. Phys.}\ }\textbf {\bibinfo {volume} {80}},\
  \bibinfo {pages} {022001} (\bibinfo {year} {2017})},\ \Eprint
  {http://arxiv.org/abs/1512.01203} {arXiv:1512.01203 [hep-th]} \BibitemShut
  {NoStop}%
\bibitem [{\citenamefont {Kusenko}\ and\ \citenamefont
  {Shaposhnikov}(1998)}]{Kusenko:1997si}%
  \BibitemOpen
  \bibfield  {author} {\bibinfo {author} {\bibfnamefont {A.}~\bibnamefont
  {Kusenko}}\ and\ \bibinfo {author} {\bibfnamefont {M.~E.}\ \bibnamefont
  {Shaposhnikov}},\ }\href {\doibase 10.1016/S0370-2693(97)01375-0} {\bibfield
  {journal} {\bibinfo  {journal} {Phys. Lett.}\ }\textbf {\bibinfo {volume}
  {B418}},\ \bibinfo {pages} {46} (\bibinfo {year} {1998})},\ \Eprint
  {http://arxiv.org/abs/hep-ph/9709492} {arXiv:hep-ph/9709492 [hep-ph]}
  \BibitemShut {NoStop}%
\bibitem [{\citenamefont {Affleck}\ and\ \citenamefont
  {Dine}(1985)}]{Affleck:1984fy}%
  \BibitemOpen
  \bibfield  {author} {\bibinfo {author} {\bibfnamefont {I.}~\bibnamefont
  {Affleck}}\ and\ \bibinfo {author} {\bibfnamefont {M.}~\bibnamefont {Dine}},\
  }\href {\doibase 10.1016/0550-3213(85)90021-5} {\bibfield  {journal}
  {\bibinfo  {journal} {Nucl. Phys.}\ }\textbf {\bibinfo {volume} {B249}},\
  \bibinfo {pages} {361} (\bibinfo {year} {1985})}\BibitemShut {NoStop}%
\bibitem [{\citenamefont {Dine}\ and\ \citenamefont
  {Kusenko}(2003)}]{Dine:2003ax}%
  \BibitemOpen
  \bibfield  {author} {\bibinfo {author} {\bibfnamefont {M.}~\bibnamefont
  {Dine}}\ and\ \bibinfo {author} {\bibfnamefont {A.}~\bibnamefont {Kusenko}},\
  }\href {\doibase 10.1103/RevModPhys.76.1} {\bibfield  {journal} {\bibinfo
  {journal} {Rev. Mod. Phys.}\ }\textbf {\bibinfo {volume} {76}},\ \bibinfo
  {pages} {1} (\bibinfo {year} {2003})},\ \Eprint
  {http://arxiv.org/abs/hep-ph/0303065} {arXiv:hep-ph/0303065 [hep-ph]}
  \BibitemShut {NoStop}%
\bibitem [{\citenamefont {Allahverdi}\ \emph {et~al.}(2010)\citenamefont
  {Allahverdi}, \citenamefont {Dutta},\ and\ \citenamefont
  {Sinha}}]{Allahverdi:2010im}%
  \BibitemOpen
  \bibfield  {author} {\bibinfo {author} {\bibfnamefont {R.}~\bibnamefont
  {Allahverdi}}, \bibinfo {author} {\bibfnamefont {B.}~\bibnamefont {Dutta}}, \
  and\ \bibinfo {author} {\bibfnamefont {K.}~\bibnamefont {Sinha}},\ }\href
  {\doibase 10.1103/PhysRevD.82.035004} {\bibfield  {journal} {\bibinfo
  {journal} {Phys. Rev.}\ }\textbf {\bibinfo {volume} {D82}},\ \bibinfo {pages}
  {035004} (\bibinfo {year} {2010})},\ \Eprint {http://arxiv.org/abs/1005.2804}
  {arXiv:1005.2804 [hep-ph]} \BibitemShut {NoStop}%
\bibitem [{\citenamefont {Kitano}\ \emph {et~al.}(2008)\citenamefont {Kitano},
  \citenamefont {Murayama},\ and\ \citenamefont {Ratz}}]{Kitano:2008tk}%
  \BibitemOpen
  \bibfield  {author} {\bibinfo {author} {\bibfnamefont {R.}~\bibnamefont
  {Kitano}}, \bibinfo {author} {\bibfnamefont {H.}~\bibnamefont {Murayama}}, \
  and\ \bibinfo {author} {\bibfnamefont {M.}~\bibnamefont {Ratz}},\ }\href
  {\doibase 10.1016/j.physletb.2008.09.049} {\bibfield  {journal} {\bibinfo
  {journal} {Phys. Lett.}\ }\textbf {\bibinfo {volume} {B669}},\ \bibinfo
  {pages} {145} (\bibinfo {year} {2008})},\ \Eprint
  {http://arxiv.org/abs/0807.4313} {arXiv:0807.4313 [hep-ph]} \BibitemShut
  {NoStop}%
\bibitem [{\citenamefont {Chen}\ and\ \citenamefont
  {Takhistov}(2019)}]{Chen:2018uzu}%
  \BibitemOpen
  \bibfield  {author} {\bibinfo {author} {\bibfnamefont {M.-C.}\ \bibnamefont
  {Chen}}\ and\ \bibinfo {author} {\bibfnamefont {V.}~\bibnamefont
  {Takhistov}},\ }\href {\doibase 10.1007/JHEP05(2019)101} {\bibfield
  {journal} {\bibinfo  {journal} {JHEP}\ }\textbf {\bibinfo {volume} {05}},\
  \bibinfo {pages} {101} (\bibinfo {year} {2019})},\ \Eprint
  {http://arxiv.org/abs/1812.09341} {arXiv:1812.09341 [hep-ph]} \BibitemShut
  {NoStop}%
\bibitem [{\citenamefont {Sugiyama}\ \emph {et~al.}(2020)\citenamefont
  {Sugiyama}, \citenamefont {Kurita},\ and\ \citenamefont
  {Takada}}]{Sugiyama:2019dgt}%
  \BibitemOpen
  \bibfield  {author} {\bibinfo {author} {\bibfnamefont {S.}~\bibnamefont
  {Sugiyama}}, \bibinfo {author} {\bibfnamefont {T.}~\bibnamefont {Kurita}}, \
  and\ \bibinfo {author} {\bibfnamefont {M.}~\bibnamefont {Takada}},\ }\href
  {\doibase 10.1093/mnras/staa407} {\bibfield  {journal} {\bibinfo  {journal}
  {Mon. Not. Roy. Astron. Soc.}\ }\textbf {\bibinfo {volume} {493}},\ \bibinfo
  {pages} {3632} (\bibinfo {year} {2020})},\ \Eprint
  {http://arxiv.org/abs/1905.06066} {arXiv:1905.06066 [astro-ph.CO]}
  \BibitemShut {NoStop}%
\bibitem [{\citenamefont {Wang}\ \emph {et~al.}(2018)\citenamefont {Wang},
  \citenamefont {Wang}, \citenamefont {Huang},\ and\ \citenamefont
  {Li}}]{Wang:2016ana}%
  \BibitemOpen
  \bibfield  {author} {\bibinfo {author} {\bibfnamefont {S.}~\bibnamefont
  {Wang}}, \bibinfo {author} {\bibfnamefont {Y.-F.}\ \bibnamefont {Wang}},
  \bibinfo {author} {\bibfnamefont {Q.-G.}\ \bibnamefont {Huang}}, \ and\
  \bibinfo {author} {\bibfnamefont {T.~G.~F.}\ \bibnamefont {Li}},\ }\href
  {\doibase 10.1103/PhysRevLett.120.191102} {\bibfield  {journal} {\bibinfo
  {journal} {Phys. Rev. Lett.}\ }\textbf {\bibinfo {volume} {120}},\ \bibinfo
  {pages} {191102} (\bibinfo {year} {2018})},\ \Eprint
  {http://arxiv.org/abs/1610.08725} {arXiv:1610.08725 [astro-ph.CO]}
  \BibitemShut {NoStop}%
\bibitem [{\citenamefont {Carr}\ \emph {et~al.}(2010)\citenamefont {Carr},
  \citenamefont {Kohri}, \citenamefont {Sendouda},\ and\ \citenamefont
  {Yokoyama}}]{Carr:2009jm}%
  \BibitemOpen
  \bibfield  {author} {\bibinfo {author} {\bibfnamefont {B.~J.}\ \bibnamefont
  {Carr}}, \bibinfo {author} {\bibfnamefont {K.}~\bibnamefont {Kohri}},
  \bibinfo {author} {\bibfnamefont {Y.}~\bibnamefont {Sendouda}}, \ and\
  \bibinfo {author} {\bibfnamefont {J.}~\bibnamefont {Yokoyama}},\ }\href
  {\doibase 10.1103/PhysRevD.81.104019} {\bibfield  {journal} {\bibinfo
  {journal} {Phys. Rev.}\ }\textbf {\bibinfo {volume} {D81}},\ \bibinfo {pages}
  {104019} (\bibinfo {year} {2010})},\ \Eprint {http://arxiv.org/abs/0912.5297}
  {arXiv:0912.5297 [astro-ph.CO]} \BibitemShut {NoStop}%
\bibitem [{\citenamefont {Dasgupta}\ \emph {et~al.}(2020)\citenamefont
  {Dasgupta}, \citenamefont {Laha},\ and\ \citenamefont
  {Ray}}]{Dasgupta:2019cae}%
  \BibitemOpen
  \bibfield  {author} {\bibinfo {author} {\bibfnamefont {B.}~\bibnamefont
  {Dasgupta}}, \bibinfo {author} {\bibfnamefont {R.}~\bibnamefont {Laha}}, \
  and\ \bibinfo {author} {\bibfnamefont {A.}~\bibnamefont {Ray}},\ }\href
  {\doibase 10.1103/PhysRevLett.125.101101} {\bibfield  {journal} {\bibinfo
  {journal} {Phys. Rev. Lett.}\ }\textbf {\bibinfo {volume} {125}},\ \bibinfo
  {pages} {101101} (\bibinfo {year} {2020})},\ \Eprint
  {http://arxiv.org/abs/1912.01014} {arXiv:1912.01014 [hep-ph]} \BibitemShut
  {NoStop}%
\bibitem [{\citenamefont {Laha}(2019)}]{Laha:2019ssq}%
  \BibitemOpen
  \bibfield  {author} {\bibinfo {author} {\bibfnamefont {R.}~\bibnamefont
  {Laha}},\ }\href {\doibase 10.1103/PhysRevLett.123.251101} {\bibfield
  {journal} {\bibinfo  {journal} {Phys. Rev. Lett.}\ }\textbf {\bibinfo
  {volume} {123}},\ \bibinfo {pages} {251101} (\bibinfo {year} {2019})},\
  \Eprint {http://arxiv.org/abs/1906.09994} {arXiv:1906.09994 [astro-ph.HE]}
  \BibitemShut {NoStop}%
\bibitem [{\citenamefont {DeRocco}\ and\ \citenamefont
  {Graham}(2019)}]{DeRocco:2019fjq}%
  \BibitemOpen
  \bibfield  {author} {\bibinfo {author} {\bibfnamefont {W.}~\bibnamefont
  {DeRocco}}\ and\ \bibinfo {author} {\bibfnamefont {P.~W.}\ \bibnamefont
  {Graham}},\ }\href {\doibase 10.1103/PhysRevLett.123.251102} {\bibfield
  {journal} {\bibinfo  {journal} {Phys. Rev. Lett.}\ }\textbf {\bibinfo
  {volume} {123}},\ \bibinfo {pages} {251102} (\bibinfo {year} {2019})},\
  \Eprint {http://arxiv.org/abs/1906.07740} {arXiv:1906.07740 [astro-ph.CO]}
  \BibitemShut {NoStop}%
\bibitem [{\citenamefont {Griest}\ \emph {et~al.}(2014)\citenamefont {Griest},
  \citenamefont {Cieplak},\ and\ \citenamefont {Lehner}}]{Griest:2013aaa}%
  \BibitemOpen
  \bibfield  {author} {\bibinfo {author} {\bibfnamefont {K.}~\bibnamefont
  {Griest}}, \bibinfo {author} {\bibfnamefont {A.~M.}\ \bibnamefont {Cieplak}},
  \ and\ \bibinfo {author} {\bibfnamefont {M.~J.}\ \bibnamefont {Lehner}},\
  }\href {\doibase 10.1088/0004-637X/786/2/158} {\bibfield  {journal} {\bibinfo
   {journal} {Astrophys. J.}\ }\textbf {\bibinfo {volume} {786}},\ \bibinfo
  {pages} {158} (\bibinfo {year} {2014})},\ \Eprint
  {http://arxiv.org/abs/1307.5798} {arXiv:1307.5798 [astro-ph.CO]} \BibitemShut
  {NoStop}%
\bibitem [{\citenamefont {Tisserand}\ \emph {et~al.}(2007)\citenamefont
  {Tisserand} \emph {et~al.}}]{Tisserand:2006zx}%
  \BibitemOpen
  \bibfield  {author} {\bibinfo {author} {\bibfnamefont {P.}~\bibnamefont
  {Tisserand}} \emph {et~al.} (\bibinfo {collaboration} {EROS-2}),\ }\href
  {\doibase 10.1051/0004-6361:20066017} {\bibfield  {journal} {\bibinfo
  {journal} {Astron. Astrophys.}\ }\textbf {\bibinfo {volume} {469}},\ \bibinfo
  {pages} {387} (\bibinfo {year} {2007})},\ \Eprint
  {http://arxiv.org/abs/astro-ph/0607207} {arXiv:astro-ph/0607207 [astro-ph]}
  \BibitemShut {NoStop}%
\bibitem [{\citenamefont {Ali-Haïmoud}\ and\ \citenamefont
  {Kamionkowski}(2017)}]{Ali-Haimoud:2016mbv}%
  \BibitemOpen
  \bibfield  {author} {\bibinfo {author} {\bibfnamefont {Y.}~\bibnamefont
  {Ali-Haïmoud}}\ and\ \bibinfo {author} {\bibfnamefont {M.}~\bibnamefont
  {Kamionkowski}},\ }\href {\doibase 10.1103/PhysRevD.95.043534} {\bibfield
  {journal} {\bibinfo  {journal} {Phys. Rev.}\ }\textbf {\bibinfo {volume}
  {D95}},\ \bibinfo {pages} {043534} (\bibinfo {year} {2017})},\ \Eprint
  {http://arxiv.org/abs/1612.05644} {arXiv:1612.05644 [astro-ph.CO]}
  \BibitemShut {NoStop}%
\bibitem [{\citenamefont {Poulin}\ \emph {et~al.}(2017)\citenamefont {Poulin},
  \citenamefont {Serpico}, \citenamefont {Calore}, \citenamefont {Clesse},\
  and\ \citenamefont {Kohri}}]{Poulin:2017bwe}%
  \BibitemOpen
  \bibfield  {author} {\bibinfo {author} {\bibfnamefont {V.}~\bibnamefont
  {Poulin}}, \bibinfo {author} {\bibfnamefont {P.~D.}\ \bibnamefont {Serpico}},
  \bibinfo {author} {\bibfnamefont {F.}~\bibnamefont {Calore}}, \bibinfo
  {author} {\bibfnamefont {S.}~\bibnamefont {Clesse}}, \ and\ \bibinfo {author}
  {\bibfnamefont {K.}~\bibnamefont {Kohri}},\ }\href {\doibase
  10.1103/PhysRevD.96.083524} {\bibfield  {journal} {\bibinfo  {journal} {Phys.
  Rev.}\ }\textbf {\bibinfo {volume} {D96}},\ \bibinfo {pages} {083524}
  (\bibinfo {year} {2017})},\ \Eprint {http://arxiv.org/abs/1707.04206}
  {arXiv:1707.04206 [astro-ph.CO]} \BibitemShut {NoStop}%
\bibitem [{\citenamefont {Niikura}\ \emph
  {et~al.}(2019{\natexlab{b}})\citenamefont {Niikura}, \citenamefont {Takada},
  \citenamefont {Yokoyama}, \citenamefont {Sumi},\ and\ \citenamefont
  {Masaki}}]{Niikura:2019kqi}%
  \BibitemOpen
  \bibfield  {author} {\bibinfo {author} {\bibfnamefont {H.}~\bibnamefont
  {Niikura}}, \bibinfo {author} {\bibfnamefont {M.}~\bibnamefont {Takada}},
  \bibinfo {author} {\bibfnamefont {S.}~\bibnamefont {Yokoyama}}, \bibinfo
  {author} {\bibfnamefont {T.}~\bibnamefont {Sumi}}, \ and\ \bibinfo {author}
  {\bibfnamefont {S.}~\bibnamefont {Masaki}},\ }\href {\doibase
  10.1103/PhysRevD.99.083503} {\bibfield  {journal} {\bibinfo  {journal} {Phys.
  Rev.}\ }\textbf {\bibinfo {volume} {D99}},\ \bibinfo {pages} {083503}
  (\bibinfo {year} {2019}{\natexlab{b}})},\ \Eprint
  {http://arxiv.org/abs/1901.07120} {arXiv:1901.07120 [astro-ph.CO]}
  \BibitemShut {NoStop}%
\bibitem [{\citenamefont {Abrams}\ and\ \citenamefont
  {Takada}(2020)}]{Abrams:2020jvs}%
  \BibitemOpen
  \bibfield  {author} {\bibinfo {author} {\bibfnamefont {N.~S.}\ \bibnamefont
  {Abrams}}\ and\ \bibinfo {author} {\bibfnamefont {M.}~\bibnamefont
  {Takada}},\ }\href@noop {} {\  (\bibinfo {year} {2020})},\ \Eprint
  {http://arxiv.org/abs/2006.05578} {arXiv:2006.05578 [astro-ph.GA]}
  \BibitemShut {NoStop}%
\end{thebibliography}%
 
\end{document}